\documentclass[useAMS,usenatbib]{stabcy}
\usepackage{amssymb,amsmath,epsfig}
\allowdisplaybreaks

\title[Electromagnetic Field and
Cylindrical Compact Objects] {Electromagnetic Field and Cylindrical
Compact Objects in Modified Gravity}
\author[Z. Yousaf and M. Zaeem-ul-Haq Bhatti]{Z.
Yousaf\thanks{zeeshan.math@pu.edu.pk (ZY)} and M. Zaeem-ul-Haq Bhatti \thanks{mzaeem.math@pu.edu.pk (MZB)}\\
Department of Mathematics, University of the Punjab,\\
Quaid-i-Azam Campus, Lahore-54590, Pakistan.}
\begin{document}

\date{}

\pagerange{\pageref{}--\pageref{}} \pubyear{}

\maketitle

\label{firstpage}

\begin{abstract}
In this paper, we have investigated the role of different fluid
parameters particularly electromagnetic field and $f(R)$ corrections
on the evolution of cylindrical compact object. We have explored the
modified field equations, kinematical quantities and dynamical
equations. An expression for the mass function has been found in
comparison with the Misner-Sharp formalism in modified gravity,
after which different mass radius diagrams are drawn. The coupled
dynamical transport equation have been formulated to discuss the
role of thermoinertial effects on the inertial mass density of the
cylindrical relativistic interior. Finally, we have presented a
framework, according to which all possible solutions of the metric
$f(R)$-Maxwell field equations coupled with static fluid can be
written through set of scalar functions. It is found that modified
gravity induced by Lagrangians $f(R)=\alpha R^2,~f(R)=\alpha
R^2-\beta R$ and $f(R)=\frac{\alpha R^2-\beta R}{1+\gamma R}$ are
likely to host more massive cylindrical compact objects with smaller
radii as compared to GR.
\end{abstract}


\section{Introduction}

It seems established from direct observations \citep{b1a,b1b,b1c} of
high red-shift supernovae as well as the cosmic microwave background
fluctuations that the universe expands with acceleration at present
state indicating the presence of dark energy. Thus, in the study of
theoretical physics, the acceleration of the universe have presented
some greatest issues. Such investigations provide motivation to
discuss the modification in gravitational dynamics at cosmological
scales. A number of attempts \citep{ya3, b2a, b2b} have been made by
many relativistic astrophysicists to describe the accelerating
cosmologies at different epochs. Particularly, a mysterious type of
energy dubbed as dark energy is involved in the explanation of the
current universe acceleration. Still, there is no suitable solid
theoretical outcomes for the source of this exotic distribution
appearing at the current epoch.

In this paper, we have focused our attention at the direct addition
of higher order curvature invariants in the Einstein-Hilbert (EH)
action. The motivation to take such a study under consideration is
that higher order curvature invariants emerge within strong gravity
regions inside the compact objects. Different authors have discussed
the modification in the EH action with terms that are pretty
effective in high-curvature regimes. \citet{1} have investigated
that cosmic acceleration can emerge with very small corrections of
the form $R^n$ in the usual gravitational action of general
relativity (GR) with $n<0$ and $R$ as the curvature scalar. The
modifications in gravity model describe the accelerated expansion
but it also contains some instabilities \citep{2}. \citet{3}
explored the instabilities in the matter due to extra curvature
ingredients in the star configuration with a specific model of
$f(R)=R-\frac{\mu^4}{R}$. \citet{4} have presented a comprehensive
review of $f(R)$ theory of gravity describing its action, field
equations, equivalence with other higher order theories, viability
criteria and astrophysical applications including all its know
formalisms namely, metric, Palatini and metric affine gravity.

The issue of gravitational collapse of massive stars has attracted
many researchers in the study of gravitational physics as it is
expected to play a vital role. \citet{5} investigated that
sufficiently strong shearing effects are responsible for the
appearance of naked singularity (instead of black hole formation)
during the collapse of a compact object. \citet{6} studied that the
isotropic and homogeneous models exist under certain conditions in
the Palatini formalism of $f(R)$ theory during the stellar
gravitational collapse. \citet{7} explored the appearance of
curvature singularity during the process of spherical collapse with
$f(R)$ environment and deduced that viable $f(R)$ gravity models
could become free of such singularities. \citet{8} studied the
collapsing phenomena within the framework of modified gravitational
theories which represents an important tool for late-time
cosmological acceleration. \citet{9} found the non-static exact
solutions for dust cloud in $f(R)$ gravity describing gravitational
collapse. Sharif and his collaborators \citep{9a3, 9a1, 9a2, 9a4,
9a5} have presented a effects of several fluid parameters in the
modeling of gravitational implosion of relativistic interiors.

Most of the stars in the universe are main sequence stars which on
evolutions collapses and leaves a compact object which is supported
by pressure gradients in the star core. Recent observations show
that many compact objects exist in nature like X-ray sources 4U
1728-34, X-ray burster 4U 1820-30, millisecond pulsar SAX J
1808.4-3658, X-ray pulsar Her X-1, PSR 0943+10 and RX J185635-3754,
whose predicted radii and masses are not consistent with standard
neutron star models. Some theoretical advances indicate that
pressure at the interior of such stars are likely to be anisotropic
as their density is greater than that of a neutron star, thus,
radial and tangential pressures exist. Recently, different authors
\citep{w8a, w8b, w8c, w8d} have explored the models of compact stars
with anisotropic fluid configurations. It is found that Fermi gas
under the influence of magnetic field generates pressure anisotropy
\citep{10a, 10b}. One can also examine the existence of anisotropy
in wormholes \citep{11} and in gravastars \citep{12a, 12b} which are
known as peculiar solutions of the field equations. \citet{13} have
formulated a framework for strongly dissipating and highly
anisotropic hydrodynamical system. They observed that the
thermalization of the system during the evolution of matter slows
down due to anisotropy by fixing initial profile of energy density
and varying initial entropy density, simultaneously.

Electromagnetic studies in curved spacetimes have a long primordial
history from the direct coupling of Einstein (or modified) and
Maxwell fields which is interpreted as the scattering of
electromagnetic waves due to spacetime curvature. \citet{14} derived
the solutions of field equations in a new gravity for a static
spherically symmetric star configuration, when the electromagnetic
effects are included in the Lagrangian. \citet{15} presented a class
of solutions for the field equations in a modified gravity theory
with negative cosmological constant and electromagnetic field which
are interpreted as black brane solutions. \citet{16} discussed the
electromagnetic field for general curved spacetime filled with
perfect fluid by using a covariant approach in GR. \citet{ya34}
explored a crucial role of electromagnetic field on the study of
some scalar functions obtained from the splitting of the curvature
tensor and named them as structure scalars. These structure scalars
are endowed with different physical meanings in the study of
relativistic astrophysics, particularly, under the influence of
electromagnetic field \citep{18}. \citet{19} found that the
electromagnetic radiations or electromagnetic waves does produce the
rotation in the system and consequently is responsible for frame
dragging. \citet{ya31} brought in the consequences of structure
scalars for in the formulation of several fundamental properties of
cylindrically symmetric relativistic interiors. \citet{ya32}
explored influence of $f(R)$ extra curvature terms on the
cylindrical gravitational implosion with the help of $f(R)$
structure scalars.

Here, a systematic construction of the scalar functions have been
presented for charged cylindrical relativistic object in the
background of $f(R)$ gravity model. Particularly, we have explored
the effects of fluid variables as well as $f(R)$ corrections on the
structure and evolution of radiating cylindrical compact object. In
this respect, set of governing equations are calculated and
expressed in terms of $f(R)$ structure scalars in order to emphasize
the importance of such scalars in the dynamical analysis of
collapsing objects. We also highlighted their significance in the
modeling of anisotropic and isotropic cylindrical collapsing system.

The paper is organized in the following fashion. After the
formulations of basic equations for cylindrical anisotropic
radiating system in section \textbf{2}. Section \textbf{3} explore
modified versions of structure scalars as well as conservations laws
framed within viable $f(R)$ model. We then formulate two important
evolution expressions that will make correspondence between Weyl
scalar, dissipating and non-dissipating fluid parameters and some
scalar variables. Section \textbf{4} is aimed to develop modified
form of cylindrical mass function as well as dynamical transport
equation. We also investigate the influence of inertial thermal
effects on the mass density during evolution of stellar interior. In
section \textbf{5}, we discuss anisotropic as well as isotropic
static cylindrical models. The results are summarized in the last
section.

\section[]{Charged Dissipative Anisotropic Relativistic Cylinders}

The notion of $f(R)$ gravity as a possible modifications in the
gravitational framework of GR received much attention of
researchers. This theory provides numerous interesting results in
the field of physics and cosmology like plausible explanation to the
accelerating cosmic expansion \citep{ya3, b2a, b2b}. The main theme
of this theory is to substitute an algebraic general function of the
Ricci scalar, $f(R)$, instead of the cosmological constant in the
standard EH action. It can be written as
\begin{equation}\label{1}
S_{f(R)}=\frac{1}{2\kappa}\int d^4x\sqrt{-g}f(R)+S_M,
\end{equation}
in which $\kappa$ is the coupling constant, $g$ is the determinant
of the tensor, $g_{\alpha\beta}$, while $S_M$ is the action for
matter density. When the curvature has a sufficiently large value,
then $f(R)$ in the above equation attains some constant value thus
behaving as a cosmological constant ($\Lambda$). However, in other
epochs, the non zero value of $f_R\equiv\frac{df}{dR}$ give rise to
complicated dynamics, thereby presenting it as a phenomenology rich
gravitational theory \citep{4}. One can be obtain $f(R)$ field
equations by varying Eq.(\ref{1}) with respect to $g_{\alpha\beta}$
and is given as follows
\begin{equation}\label{2}
R_{\alpha\beta}f_R-\frac{1}{2}f(R)g_{\alpha\beta}+\left(g_{\alpha\beta}{\Box}
-\nabla_{\alpha}\nabla_{\beta}\right)f_R={\kappa}T_{\alpha\beta},
\end{equation}
where $R_{\alpha\beta}$ and $T_{\alpha\beta}$ are Ricci tensor and
standard energy momentum tensor, respectively. The above field
equation is a $4th$ order differential equation due to the existence
of second order derivatives of $f_R=f_R(R)$ (It is worthy to mention
that $f_R$ contains further second order derivatives of the metric
functions). Some researchers have dubbed $f_R$ as scalaron that
indicates new scalar degree of freedom. Its equation of motion can
be specified by taking the trace of Eq.(\ref{2}) as
\begin{align}\label{3}
\Box{f_R}=\frac{1}{3}\left(2f+R+{\kappa}T-Rf_R\right),
\end{align}
where $T\equiv T^\beta_{~\beta}$. On setting, $f_R\rightarrow0$,
$f(R)\rightarrow\Lambda$, one can get a well-known result, i.e.,
$R=-(\kappa{T}+2\Lambda)$. In terms of Einstein tensor,
$G_{\alpha\beta}$, Eq.(\ref{2}) can be written as
\begin{equation}\label{4}
G_{\alpha\beta}=\frac{\kappa}{f_R}(\overset{(D)}
{T_{\alpha\beta}}+T_{\alpha\beta}),
\end{equation}
where
\begin{equation*}
\overset{(D)}{T_{\alpha\beta}}=\frac{1}{\kappa}\left\{
\nabla_{\alpha}\nabla_{\beta}f_R-\Box
f_Rg_{\alpha\beta}+(f-Rf_R)\frac{g_{\alpha\beta}}{2}\right\},
\end{equation*}
is an effective $f(R)$ gravitational interaction.

We consider a non-rotating diagonal cylindrical relativistic system
characterized by the following line element \citep{ya31}
\begin{equation}\label{5}
ds^2=-A^2(dt^{2}-dr^{2})+B^2dz^2 +C^2d\phi^2,
\end{equation}
filled locally anisotropic matter configuration that is dissipating
in the mode of heat radiations. Here $A,~B$ and $C$ are the
functions of $t$ and $r$. The mathematical formula describing fluid
distribution within the cylindrical relativistic celestial object is
\begin{align}\nonumber
&T_{\alpha\beta}=(\mu+P)V_\alpha
V_\beta+(P_\phi-P_r)\left(K_{\alpha}K_\beta-\frac{h_{\alpha\beta}}{3}\right)
+(P_z-P_r)\\\label{6}
&\times\left(S_{\alpha}S_\beta-\frac{h_{\alpha\beta}}{3}\right)+q_\alpha
V_\beta+Pg_{\alpha\beta}+q_\beta V_\alpha.
\end{align}
The evolution of this fluid distribution is characterized by
timelike $3D$ boundary, $\Sigma$. Here $\mu$ is a matter energy
density, $P_r,~P_z$ and $P_\phi$ are the corresponding principal
stresses, while $q_\alpha$ is a four vector radiating heat along the
radial direction. The unitary vectors,
$S_\alpha,~K_\alpha,~V_\alpha$ and $L_\alpha$, are configuring with
the fact that they determine canonical orthonormal tetrad. The two
vectors of these tetrads, $S_\alpha,~K_\alpha$, are tangent to the
orbits of $2D$ group, $L_\alpha$ is a boundary orthogonal to
$V_\alpha$ and to these orbits. These four vectors under comoving
relative motion are defined as
$S_\alpha=B\delta^2_{~\beta},~V_\alpha=-A\delta^1_{~\beta},~K_\alpha=C\delta^3_{~\beta}$
and $L_\alpha=A\delta^1_{~\beta}$ alongwith the following
constraints
\begin{align}\nonumber
&S^\alpha S_\alpha=L^\alpha L_\alpha=K_\alpha
K^\alpha=1,~V^{\alpha}V_{\alpha}=-1,\\\nonumber &V^\alpha
S_\alpha=V^\alpha L_\alpha=K_\alpha V^\alpha=K_\alpha S^\alpha=0.
\end{align}

If the matter configuration within the cylindrical system is charged
then, the tensorial formulation describing electromagnetic effects
is given by
\begin{equation}\label{7}
E_{\alpha\beta}=\frac{1}{4\pi}\left(-F^{\gamma}_{\alpha}F_{\beta\gamma}
+\frac{1}{4}F^{\gamma\delta}F_{\gamma\delta}g_{\alpha\beta}\right),
\end{equation}
in which $F_{\alpha\beta}$ is a Maxwell field tensor defied in terms
of four potential, $\phi_\alpha$, as
$F_{\alpha\beta}=\phi_{\beta,\alpha}-\phi_{\alpha,\beta}$. The
corresponding Maxwell equations of motion are
\begin{equation}\label{8}
F^{\alpha\beta}_{~~;\beta}={\mu}_{0}J^{\alpha},\quad
F_{[\alpha\beta;\gamma]}=0,
\end{equation}
in which $J^\alpha$ is a four current, while $\mu_0$ indicates
magnetic permeability. Here we suppose that, due to comoving
relativistic motion of the fluid, the electric charge within the
matter is at rest which eventually give rise to zero magnetic field.
Thus we have
\begin{equation*}\nonumber
\phi_{\alpha}={\phi}(t,r){\delta^{0}_{\alpha}},\quad
J^{\alpha}={\rho}(t,r)V^{\alpha},
\end{equation*}
where $\phi$ represents scalar potential and $\rho$ indicates the
charge density. Equation (\ref{8}) provide the following equation of
motions for charged relativistic cylindrical systems
\begin{eqnarray}\label{9}
&&\frac{\partial^2{\phi}}{{\partial}r^2}-\left(\frac{2A'}{A}-\frac{B'}{B}
-\frac{C'}{C}\right)\frac{\partial\phi}{{\partial}r}
=4\pi{\rho}A^3,\\\label{10} &&\frac{\partial}{\partial
t}\left(\frac{\partial\phi}{{\partial}r}\right)
-\left(\frac{2\dot{A}}{A}-\frac{\dot{B}}{B}-\frac{\dot{C}}{C}\right)\frac
{\partial\phi}{{\partial}r}=0,
\end{eqnarray}
where over dot represents $\frac{\partial}{\partial{t}}$ while prime
indicates $\frac{\partial}{r}$. The solution of Eq.(\ref{9})
provides
\begin{equation*}
\phi'=\frac{A^2s(r)}{BC},
\end{equation*}
where
\begin{equation}\label{11}
s(r)=2\pi\int^r_{0}{\rho}{ABC}dr,
\end{equation}
is the total charge within cylindrical system of radius $r$. This
quantity can also be evaluated via conservation law of four current,
i.e., $J^\mu_{~;\mu}=0.$ The electric field strength/intensity that
describes charge per unit cylindrical areal surface is written as
\begin{equation}\label{12}
E(t,r)=\frac{s}{2{\pi}B}.
\end{equation}

The $f(R)$ field equations (\ref{4}) for cylindrically symmetric
system coupled with dissipative matter configuration with charged
background turn out to be
\begin{align}\nonumber
&\frac{\dot{C}\dot{B}}{BC}-\frac{C''}{C}-\frac{B''}
{B}-\frac{B'C'}{BC}+\alpha_1=\frac{\kappa{A^2}}{f_R}\left[\mu+\frac{2\pi
E^2}{B^2}+\frac{1}{\kappa}\right.\\\label{13} &\times\left.\left
\{\frac{f''_R}{A^2}-\frac{\dot{A}\dot{f_R}}{A^3}-\frac{A'f'_R}{A^3}
-\gamma_1\right\}\right],\\\nonumber
&\left(\frac{B'}{B}+\frac{C'}{C}\right)\frac{\dot{A}}{A}-\frac{
\dot{C'}}{C}-\frac{\dot{B'}}{B}+\left(\frac{\dot{B}}{B}+\frac{\dot{C}}{C}\right)
\frac{A'}{A}=\frac{\kappa}{f_R}\left[-qA^2 \right.\\\label{14}
&\left.+\frac{1}{\kappa}\left\{\dot{f'_R}-\frac{A'}{A}\dot{f_R}-\frac{\dot{A}}{A}f'_R\right\}\right],\\\nonumber
&\frac{B'{C'}}{BC}-\frac{\ddot{B}}{B}-\frac{\dot{B}\dot{C}}{BC}
-\frac{\ddot{C}}{C}+\alpha_1=\frac{\kappa{A^2}}{f_R}\left[P_r-\frac{2\pi
E^2}{B^2}+\frac{ 1}{\kappa}\right.\\\label{15}
&\times\left.\left\{\frac{\ddot{f_R}}{A^2}-\frac{A'f'_R}{A^3}-\frac{\dot{A}\dot{f_R}}
{A^3}+\gamma_1\right\}\right],\\\nonumber
&\left(\frac{B}{A}\right)^2\left[\beta_1
+\frac{C''}{C}-\frac{\ddot{C}}{C}\right]=\frac{\kappa{B^2}}{f_R}\left[P_z+\frac{2\pi
E^2}{B^2}+\frac{1} {\kappa}\left\{\delta_1\right.\right.\\\label{16}
&\left.\left.-\frac{C'}{C}\frac{f'_R}{A^2}\right\}\right],\\\nonumber
&\left(\frac{C}{A}\right)^2\left[\beta_1+\frac{B''}{B}-\frac{\ddot{B}}{B}
\right]=\frac{\kappa{C^2}}{f_R}\left[P_\phi+\frac{2\pi
E^2}{B^2}+\frac{1}{\kappa} \left\{\delta_1\right.\right.\\\label{17}
&\left.\left.-\frac{B'}{B}\frac{f'_R}{A^2}\right\}\right],
\end{align}
where
\begin{align}\nonumber
&\alpha_1=\left(\frac{\dot{C}}
{C}+\frac{\dot{B}}{B}\right)\frac{\dot{A}}{A}+\left(\frac{{B'}}{B}+\frac{{C'}}{C}\right)\frac{{A'}}{A}
,\\\nonumber
&\beta_1=\frac{\dot{A}^2}{A^2}-\frac{A'^2}{A^2}-\frac{\ddot{A}}{A}
+\frac{A''}{A},\\\nonumber
&\gamma_1=\frac{R}{2}\left(\frac{f}{R}-f_R\right)+\left(\frac{\dot{B}}{B}+\frac{\dot{C}}{C}\right)
\frac{\dot{f_R}}{A^2}-\left(\frac{B'}{B}+\frac{C'}{C}\right)
\frac{f'_R}{A^2},\\\nonumber
&\delta_1=\frac{R}{2}\left(\frac{f}{R}-f_R\right)+\frac{1}{A^2}
\left(\ddot{f_R}+\frac{\dot{C}}{C}\dot{f_R}-f''_R \right).
\end{align}

\section[]{Modified Structure Scalars}

In this section, we shall first notice the effects of dark source
$f(R)$ corrections in the expressions of well-known structure
scalars. We also consider well-consistent $f(R)$ model and then
discuss evolution of these function variables in the dynamics of
cylindrically symmetric relativistic systems. \citet{b24} provided
the key notion of orthogonal splitting of the Riemann tensor. Later
on, \citet{ya37} generalized this concept and discussed many
problems related to evolution of compact objects. It is seen that
Riemann curvature tensor upon orthogonal decomposition yields set of
three tensorial quantities, i.e., $X_{\alpha\beta},~Y_{\alpha\beta}$
and $Z_{\alpha\beta}$. The explicit expressions of $Y_{\alpha\beta}$
and $X_{\alpha\beta}$ are carried out by means of trace and
traceless components (known as structure scalars) as follows
\citep{ya341, ya34}
\begin{align}\label{21}
Y_{\alpha\beta}&=\left(S_\alpha
S_\beta-\frac{h_{\alpha\beta}}{3}\right)Y_S+\left(K_{\alpha}K_\beta-\frac{h_{\alpha\beta}}
{3}\right)Y_K+\frac{h_{\alpha\beta}}{3}Y_{T},\\\label{22}
X_{\alpha\beta}&=\left(S_\alpha
S_\beta-\frac{h_{\alpha\beta}}{3}\right)X_S+\left(K_{\alpha}K_\beta-\frac{h_{\alpha\beta}}
{3}\right)X_K+\frac{h_{\alpha\beta}}{3}X_{T},
\end{align}
in which $Y_T,~Y_S,~Y_K,~X_S,~X_K$ and $X_T$ are the combinations of
the corresponding quantities with respect to standard, charged and
effective dark matter configurations. These scalars can be expressed
as
\begin{align}\label{23}
&Y_T=\sum_{i=1}^3Y_T^{(i)}=T_T^{(G)}+Y_T^{(C)}+Y_T^{(D)},\\\label{24}
&Y_S=\sum_{i=1}^3Y_S^{(i)}=T_S^{(G)}+Y_S^{(C)}+Y_S^{(D)}
,\\\label{25}
&Y_K=\sum_{i=1}^3Y_K^{(i)}=T_K^{(G)}+Y_K^{(C)}+Y_K^{(D)},\\\label{26}
&X_S=\sum_{i=1}^3X_S^{(i)}=X_S^{(G)}+X_S^{(C)}+X_S^{(D)},\\\label{27}
&X_K=\sum_{i=1}^3X_K^{(i)}=X_K^{(G)}+X_K^{(C)}+X_K^{(D)},\\\label{28}
&X_T=\sum_{i=1}^3X_T^{(i)}=X_T^{(G)}+Y_K^{(C)}+X_T^{(D)}.
\end{align}
It has been observed that the magnetic part of the Weyl tensor does
not vanish for cylindrically symmetric celestial objects. This gives
rise to the birth of another tensorial term $Z_{\alpha\beta}$
\cite{ya34} (after orthogonal decomposition), whose expression is
given as
\begin{align}\label{29}
Z_{\alpha\beta}&=H_{\alpha\beta}+\frac{\kappa}{2f_R}\left(q-\frac{\overset{
~~~(D)}{T_{01}}}{A^2}\right)^\rho\epsilon_{\alpha\beta\rho},
\end{align}
From this, couple of scalar variables can be found as follows
\begin{align}\label{30}
&Z_q=\sum_{i=1}^3Z_q^{(i)}=Z_q^{(G)}+Z_q^{(C)}+Z_q^{(D)},\quad
Z_H=2H
\end{align}
By making use of Eqs.(\ref{13})-(\ref{17}) and (\ref{20}), above
mentioned modified structure scalars are found as
\begin{align}\label{31}
&Y_T^{(G)}=\frac{\kappa}{2f_R}\left(\mu+P_z+P_r+P_\phi\right), \quad
Y_T^{(D)}=\frac{2\kappa\pi{E^2}}{B^2f_R},\\\label{32} &
Y_T^{(D)}=\frac{\kappa}{2f_R}\left[\frac{1}{A^2}\left(\overset
{~~~(D)}{T_{00}}+\overset{~~~(D)}{T_{11}}\right)+\frac{
\overset{~~~(D)}{T_{22}}}{B^2}+\frac{\overset{~~~(D)}{T_{33}}}{C^2}\right]
,\\\label{33}
&Y_S^{(G)}=\mathbb{E}_S-\frac{\kappa}{2f_R}\left(P_z-P_r\right),\\\nonumber
& Y_S^{(C)}=-\frac{2\kappa\pi{E^2}}{B^2f_R},\\\label{34}
&Y_S^{(D)}=\frac{\kappa}{2f_R}\left(\frac{\overset{~~~(D)}{T_{11}}}{A^2}-\frac{\overset{~~~(D)}
{T_{22}}}{B^2}\right),~
Y_K^{(G)}=\mathbb{E}_K-\frac{\kappa}{2f_R}\left(P_\phi-P_r\right)\\\label{35}
& Y_K^{(C)}=-\frac{2\pi\kappa}{B^2f_R}E^2,\quad
Y_K^{(D)}=\frac{\kappa}{2f_R}\left(\frac{\overset{~~~(D)}{T_{11}}}{A^2}-\frac{\overset{~~~(D)}
{T_{33}}}{C^2}\right),\\\label{36}
&X_S^{(G)}=-\mathbb{E}_S-\frac{\kappa}{2f_R}\left(P_z-P_r\right),\quad
X_S^{(C)}=-\frac{2\pi\kappa}{B^2f_R}E^2,\\\label{37}
&X_S^{(D)}=\frac{\kappa}{2f_R}\left(\frac{
\overset{~~~(D)}{T_{11}}}{A^2}-\frac{\overset{~~~(D)}
{T_{22}}}{B^2}\right),~
X_K^{(G)}=-\mathbb{E}_K-\frac{\kappa}{2f_R}\left(P_\phi-P_r\right)\\\label{38}
& X_K^{(C)}=-\frac{2\pi\kappa}{B^2f_R}E^2,\quad
X_K^{(D)}=-\frac{\kappa}{2f_R}\left(\frac{\overset{~~~(D)}
{T_{33}}}{C^2}-\frac{\overset{~~~(D)}{T_{11}}}{A^2}\right),\\\label{39}
&X_T^{(G)} =\frac{\kappa}{f_R}(\mu),~
X_T^{(C)}=\frac{2\pi\kappa}{B^2f_R}E^2,~
X_T^{(D)}=\frac{\kappa\overset{~~~(D)}{T_{00}}}{2A^2f_R},\\\label{40}
&Z_q^{(G)} =\frac{\kappa{q}}{2f_R},\quad Z_q^{(C)}=0,~
Z_q^{(D)}=-\frac{\kappa\overset{~~~(D)}{T_{01}}}{2A^2f_R},~ Z_H=2H,
\end{align}
where $\mathbb{E}_S,~\mathbb{E}_K$ and $H$ are scalar corresponding
to electric and magnetic components of the Weyl tensor (for details
see \citep{ya31}). Their values are
\begin{align}\label{20}
\mathbb{E}_S&=\frac{1}{A^2}\left(\frac{1}{B^2}C_{0202}-\frac{1}{A^2}C_{0101}\right),\\\nonumber
\mathbb{E}_K&=\frac{1}{A^2}\left(\frac{1}{C^2}C_{0303}-\frac{1}{A^2}C_{0101}\right),~H=-\frac{1}{C^2A^2}C_{0313}.
\end{align}
The explicit equations for the Weyl tensor componenets
$C_{0202},~C_{0101}$, $C_{0303},~C_{0313}$ can easily be exhibit
with the help of \verb"MATLAB". It is observed from the above
formulations that in the modeling of non-static non-rotating
diagonal cylindrical systems, one needs to establish eight distinct
scalar variables. These scalars are further are further being
demarcated into three portions based on the nature of fluid
distribution. It is worthy to mention that these quantities are
evaluated by considering general formalism of metric $f(R)$
corrections. One can get GR scalar variables after application of
$f(R)=R$ in the above equations.

The conservation laws for standard, charged and effective
energy-momentum tensors yield
\begin{align}\label{41}
&\mu^*+(\mu+P_r)\Theta+q^\alpha{a_{\alpha}}+q^\alpha_{~;\alpha}
+\Pi^{\alpha\beta}\sigma_{\alpha\beta}+\frac{1}{3}\Pi^\alpha_
{~\alpha}\Theta+D_0=0,\\\nonumber
&h^{\alpha\beta}(\Pi^\mu_{~\beta;\mu}+P_{r;\beta}+q^*_\beta)
+a^\alpha(\mu+P_r)+\frac{4}{3}q^\alpha{\Theta}-\frac{4\pi
E}{A^2B^2C}\\\label{42} &\times (CE'+EC')+q^\mu\sigma^\alpha
_{~\mu}+D_1=0.
\end{align}
where subscript $*$ shows that respective quantity is evaluated
under an operator given by $g^*=g_{,\alpha}V^\alpha$, $D_0$ and
$D_1$ are $f(R)$ dark source terms mentioned in Appendix \textbf{A},
while $\Theta\equiv V^\beta_{~;\beta}$ is an expansion scalar given
by
\begin{align}\nonumber
\Theta=\left(\frac{\dot{B}}{B}+\frac{\dot{A}}{A}+\frac{\dot{C}}{C}\right)\frac{1}{A}.
\end{align}
Equation (\ref{42}) can be recasted as
\begin{align}\nonumber
&P^\dag_r+a(\mu+P_r)+q^*-\frac{1}{A}\left[(P_z-P_r)\frac{B'}
{B}\right.+(P_\phi-P_r)\left.\frac{C'}{C}\right]\\\label{43}
&-\frac{4\pi
E}{A^2B^2C}(CE'+EC')-\frac{q}{3}(\sigma_S-4\Theta+\sigma_K)+D_1=0,
\end{align}
where subscript $\dag$ indicates $g^\dag=g_{,\alpha}L^\alpha$
operator, while $\sigma_K$ and $\sigma_S$ are shear scalars, whose
values are
\begin{align}\nonumber
\sigma_K=\left(\frac{\dot{C}}{C}-\frac{\dot{A}}{A}\right)\frac{1}{A},\quad
\sigma_S=\left(\frac{\dot{B}}{B}-\frac{\dot{A}}{A}\right)\frac{1}{A}.
\end{align}

In order to present $f(R)$ theory of gravity as an acceptable
theory, one should consider a viable as well as well-consistent
$f(R)$ model. It was one of the challenging tasks of relativistic
astrophysicists to introduce a  model that can not only explain dark
energy epochs but also discuss inflationary eras of cosmos, with the
single formulation. To generate such kind of results from $f(R)$
gravity models, they must obey certain restrictions imposed by
terrestrial and solar system experiments with relativistic
background \citep{ya35}. Here, we consider a well-known realistic
$f(R)$ model that tends to unify inflation and dark energy eras. Its
mathematical formulation is given as follows
\begin{align}\label{44}
f(R)=\frac{{\alpha}R^{2n}-\beta{R^n}}{1+\gamma{R^n}}.
\end{align}
where $n$ is a non-negative and non-zero integer while
$\alpha,~\beta$ and $\gamma$ are positive constants. It is found
that this model produces instabilities-free $f(R)$ contribution in
the explanation of cosmological evolution. It yields several results
consistent with the outcomes noticed in cosmological and solar
system tests. Thus, this may be considered as a viable model in
which late-time cosmic acceleration as well as early-time
inflationary eras are naturally unified within single model.

The two very important expressions that would be helpful in the
description of evolutionary phases of radiating anisotropic
cylindrically symmetric systems can be obtained after using
Eqs.(\ref{13})-(\ref{17}), (\ref{23})-(\ref{28}) and (\ref{30}).
These expressions were firstly computed by \citet{ya31} in GR.
These, in Maxwell-$f(R)$ gravity, turn out to be
\begin{align}\nonumber
&\frac{\kappa(1+\gamma{R^n})^2}{nR^{n-1}[(1+\gamma{R^n})(2\alpha{R}^n-\beta)-\gamma{R^n}(\alpha
R^n-\beta)]}\left(2\mu+\frac{2\pi{E^2}}{B^2}\right.\\\nonumber
&+\left.\frac{n\gamma{R}^{2n}(\alpha{R^n}-\beta)}
{2(1+\gamma{R^n})^2}+\frac{{\alpha}R^{2n}(1-2n)-{\beta}R^n(1-n)}
{2(1+\gamma{R^n})} +P_z\right.\\\nonumber
&\left.+P_r+P_\phi+\frac{2\psi_{tt}}{A^2}+\frac{2\psi_{rr}}{A^2}+\frac{\psi_{zz}}{B^2}
+\frac{\psi_{\phi\phi}}{C^2}\right)^
\dag+\left(\mu+P_r+\frac{\psi_{tt}}{A^2}\right.\\\nonumber
&\left.+\frac{\psi_{rr}}{A^2}\right)\frac{3a\kappa
(1+\gamma{R^n})^2}{nR^{n-1}[(1+\gamma{R^n})(2\alpha{R}^n-\beta)-\gamma{R^n}(\alpha
R^n-\beta)]}\\\nonumber
&-\frac{2\kappa(1+\gamma{R^n})^2}{nR^{n-1}[(1+\gamma{R^n})(2\alpha{R}^n-\beta)-\gamma{R^n}(\alpha
R^n-\beta)]}\left(q-\frac{\psi_{tr}}{A^2}\right)\\\nonumber
&\times(\sigma_S-\Theta+\sigma_K)+\frac{3\kappa(1+\gamma{R^n})^2}{nR^{n-1}[(1+\gamma{R^n})(2\alpha{R}^n-\beta)-\gamma{R^n}(\alpha
R^n-\beta)]}\\\nonumber
&\times\left(q-\frac{\psi_{tr}}{A^2}\right)^*
=3(X_K-Y_K)\frac{C'}{AC}-(Y_S+Y_K-X_S-X_K)^\dag\\\label{45}
&+3(X_S-Y_S)\frac{B'}{AB}
-6H(\sigma_S-\sigma_K),\\\nonumber\\\nonumber
&-\frac{\kappa(1+\gamma{R^n})^2}{nR^{n-1}[(1+\gamma{R^n})(2\alpha{R}^n-\beta)-\gamma{R^n}(\alpha
R^n-\beta)]}\left(\mu-P_r\right.\\\nonumber
&+\frac{6\pi{E^2}}{B^2}-P_z-\frac{{\alpha}R^{2n}(1-2n)-{\beta}R^n(1-n)}{2(1+\gamma{R^n})}+\frac{n\gamma{R}^{2n}(\alpha{R^n}-\beta)}
{2(1+\gamma{R^n})^2}\\\nonumber
&\left.+2P_\phi+\frac{\psi_{tt}}{A^2}-\frac{\psi_{rr}}{A^2}-\frac{\psi_{zz}}{B^2}
+\frac{2\psi_{\phi\phi}}{C^2}\right)^\dag-\left(q-\frac{\psi_{tr}}{A^2}\right)\\\nonumber
&\frac{\kappa(1+\gamma{R^n})^2(\sigma_S
-\Theta-2\sigma_K)}{nR^{n-1}[(1+\gamma{R^n})(2\alpha{R}^n-\beta)-\gamma{R^n}(\alpha
R^n-\beta)]}\\\nonumber &
-\frac{3\kappa(1+\gamma{R^n})^2}{nR^{n-1}[(1+\gamma{R^n})(2\alpha{R}^n-\beta)-\gamma{R^n}(\alpha
R^n-\beta)]}\\\nonumber
&\times\left(P_\phi-P_r-\frac{\psi_{rr}}{A^2}
+\frac{\psi_{\phi\phi}}{C^2}\right)\frac{C'}{AC}=6H^*+(2Y_S-2X_S-Y_K\\\nonumber
&+X_K)^\dag+3(Y_S-X_S)\frac{B'}{AB}+3a(X_K-X_S-Y_K+Y_S)\\\label{46}
&+6H(\Theta-\sigma_K),
\end{align}
in which $a$ is a scalar term expressed in terms of four
acceleration as $a_\beta=aL_\beta$. Its value is $a=\frac{A'}{A^2}$.
The quantities $\psi_{\alpha\beta}$ are $f(R)$ corrections and can
easily be evaluated after using Eq.(\ref{44}) in the following
equations
\begin{align}\nonumber
\psi_{\alpha\beta}=\overset{(D)}{T_{\alpha\beta}}-\frac{1}{2}(f-Rf_R)g_{\alpha\beta}.
\end{align}

The evolution and stability of relativistic self-gravitating systems
during gravitational collapse have become interesting phenomenon not
only in GR but also in modified gravity theories. The utmost
relevance of collapsing mechanism lies at the center of stellar
structure formation which happens when smooth initial matter
configurations will ultimately collapse. This process gives rise to
overwhelm structures such as stars, stellar groups and planets. A
system begins collapsing once it experiences an inhomogeneous
stellar state. Which factors are responsible for the emergence of
inhomogeneity phases in the initially homogeneous system? To answer
this, we have explored above two relations. \cite{12} claimed that
system's inhomogeneity factors are very important in the discussion
of its collapsing behavior. So, they explored irregularities in the
energy density of spherical relativistic stars by means of Weyl
scalar. In order to see how above relations work, we consider
cylindrical isotropic non-radiating system (simplest case). Under
this scenario, Eqs.(\ref{24})-(\ref{27}) give
\begin{align}\nonumber
&Y_S=\mathbb{E}_S+\frac{\kappa}{2f_R}\left(\frac{\overset{~~~(D)}
{T_{11}}}{A^2}-\frac{4\pi{E^2}}{B^2}-\frac{\overset{~~~(D)}
{T_{22}}}{B^2}\right),\\\nonumber
&Y_K=\mathbb{E}_K+\frac{\kappa}{2f_R}\left(
\frac{\overset{~~~(D)}{T_{11}}}{A^2}-\frac{4\pi}{B^2}E^2-\frac{\overset{~~~(D)}
{T_{33}}}{C^2}\right),\\\nonumber
&X_S=-\mathbb{E}_S+\frac{\kappa}{2f_R}\left(\frac{
\overset{~~~(D)}{T_{11}}}{A^2}-\frac{4\pi}{B^2}E^2-\frac{\overset{~~~(D)}
{T_{22}}}{B^2}\right),\\\nonumber
&X_K=-\mathbb{E}_K+\frac{\kappa}{2f_R}\left(\frac{\overset{~~~(D)}
{T_{33}}}{C^2}-\frac{4\pi}{B^2}E^2-\frac{\overset{~~~(D)}{T_{11}}}{A^2}\right),
\end{align}
Using these relations along with Eq.(\ref{44}) in Eq.(\ref{46}), we
get
\begin{align}\nonumber
&-\frac{\kappa(1+\gamma{R^n})^2}{nR^{n-1}[(1+\gamma{R^n})(2\alpha{R}^n-\beta)-\gamma{R^n}(\alpha
R^n-\beta)]}\left(\mu\right.\\\nonumber
&+\frac{6\pi{E^2}}{B^2}-\frac{{\alpha}R^{2n}(1-2n)-{\beta}R^n(1-n)}
{2(1+\gamma{R^n})}+\frac{n\gamma{R}^{2n}(\alpha{R^n}-\beta)}
{2(1+\gamma{R^n})^2}\\\nonumber
&\left.+\frac{\psi_{tt}}{A^2}-\frac{\psi_{rr}}{A^2}-\frac{\psi_{zz}}{B^2}
+\frac{2\psi_{\phi\phi}}{C^2}\right)^\dag-\left(\frac{\psi_{\phi\phi}}{C^2}-\frac{\psi_{rr}}{A^2}
\right)\frac{C'}{AC}\\\nonumber &
\times\frac{3\kappa(1+\gamma{R^n})^2}{nR^{n-1}[(1+\gamma{R^n})(2\alpha{R}^n-\beta)-\gamma{R^n}(\alpha
R^n-\beta)]}\\\nonumber
&=6H^*-2\left\{\left(\frac{\psi_{zz}}{C^2}-\frac{\psi_{rr}}{A^2}+2\mathbb{E}_S\right.\right)\\\nonumber
\nonumber
&\left.\times\frac{\kappa(1+\gamma{R^n})^2}{nR^{n-1}[(1+\gamma{R^n})(2\alpha{R}^n-\beta)-\gamma{R^n}(\alpha
R^n-\beta)]}\right\}^\dag\\\nonumber
&+6\mathbb{E}_S\frac{B'}{AB}+6H(\Theta-\sigma_K)
-6a\left(\frac{\psi_{zz}}{C^2}-\frac{\psi_{rr}}{A^2}+2\mathbb{E}_S\right)\\\label{46a}
&
\times\frac{\kappa(1+\gamma{R^n})^2}{nR^{n-1}[(1+\gamma{R^n})(2\alpha{R}^n-\beta)-\gamma{R^n}(\alpha
R^n-\beta)]},
\end{align}
For conformally flat solutions. we have $\mathbb{E}_S=0=H$, then
$\mu^\dag$ is directly related with electromagnetic field and $f(R)$
dark source terms. In GR, above equation reduces to $$\mu^\dag=0
\Rightarrow \mu=\mu(t).$$ This shows that in GR, it is easy for the
system to enter or leave the homogeneous phase. However, in
Maxwell-$f(R)$ gravity, electric charge as well as $f(R)$ extra
curvature terms tend to produce hindrances for the system to leave
and enter homogeneous state. Thus, it is relatively difficult for
the system to enter in the unstable phases in Maxwell-$f(R)$ field,
thereby indicating that more stable cylindrical configurations of
stellar systems exists in modified gravity. Similar result can be
deduced from Eq.(\ref{45}).

In the framework of $f(R)$ model given in Eq.(\ref{44}), modified
versions of scalar quantities are also evaluated and are mentioned
in Appendix \textbf{A}. We know that these scalar variables are
being explored after orthogonal decomposition of Reimann curvature
tensors. It has been analyzed from the work of many researchers that
structure scalars have a crucial importance in discussing the
dynamical evolution of self-gravitating spherical celestial bodies
\citep{ya34, ya31,9a6, 9a7, ya32}. We shall now discuss the
importance of these variables in the modeling of relativistic
cylindrical stellar object in the coming sections.

\section[]{$f(R)$ Cylindrical Generalization to Spherical Misner-Sharp Mass Function}

This section is aimed to investigate thermoinertial effects on the
modified effective inertial mass of locally anisotropic non-rotating
cylindrical celestial systems in the presence of electromagnetic
field. We also discuss the influence of heat radiating parameter in
the subsequent evolution of collapsing stellar objects in the Jordan
frame of metric $f(R)$ gravity. For this, we first evaluate modified
version of dynamical-transport equation and then formulate mass
function for charged cylindrical relativistic interior with
inflationary and late time cosmic accelerating $f(R)$ model. We
shall express these terminologies in terms of modified versions of
structure scalars. The velocity of the collapsing fluid is defined
as the change in cylindrical areal radius corresponding to proper
time. This velocity is often taken to be less than zero for
collapsing relativistic interior and is given as
\begin{align}\label{47}
U=\frac{\dot{C}}{A}=C^*<0~\textmd{(for collapsing system)},
\end{align}
which after using 11 field equation gives
\begin{align*}\nonumber
&U^*=\frac{aC'}{A}-\frac{\kappa
C(1+\gamma{R^n})^2}{nR^{n-1}[(1+\gamma{R^n})(2\alpha{R}^n-\beta)-\gamma{R^n}(\alpha
R^n-\beta)]}\\\nonumber
&\times\left(P_r-\frac{2\pi{E^2}}{B^2}+\frac{\psi_{rr}}{A^2}\right)-\frac{{\alpha}R^{2n}(1-2n)
-{\beta}R^n(1-n)}{2(1+\gamma{R^n})}\\\nonumber
&+\frac{n\gamma{R}^{2n}(\alpha{R^n}-\beta)}
{2(1+\gamma{R^n})^2}-\frac{C}{A^2}\left[\frac{\ddot{B}}{B}
+\frac{\dot{B}}{B}\left(\frac{\dot{C}}{C}-\frac{\dot{A}}{A}\right)
-\frac{{B'}}{B}\right.\\\nonumber
 &\times\left.\left(\frac{C'}{C}+\frac{A'}{A}\right)\right].
\end{align*}
This equation can also be manipulated as
\begin{align}\nonumber
U^*&=-\frac{\kappa
C(1+\gamma{R^n})^2}{nR^{n-1}[(1+\gamma{R^n})(2\alpha{R}^n-\beta)-\gamma{R^n}(\alpha
R^n-\beta)]}\\\nonumber
&\times\left(P_r-\frac{2\pi{E^2}}{B^2}\right.\left.+\frac{\psi_{rr}}{A^2}\right)+\frac{aC'}{A}-\frac{{\alpha}R^{2n}(1-2n)
-{\beta}R^n(1-n)}{2(1+\gamma{R^n})}\\\label{48}&
+\frac{n\gamma{R}^{2n}(\alpha{R^n}-\beta)}
{2(1+\gamma{R^n})^2}-\frac{C}{B^2} \left(\frac{R_{2323}}{C^2}
-\frac{R_{0202}}{A^2}\right),
\end{align}
where $R_{2323}$ and $R_{0202}$ are components of curvature tensor
and are written down as
\begin{align}\nonumber
R_{2323}&=B^2\left(\frac{A'B'}{AB}+\frac{\dot{A}\dot{B}}{AB}-\frac{\ddot{B}}{B}\right),\\\nonumber
R_{0202}&=-\frac{C^2}{A^2}\left(\frac{B'C'}{BC}-\frac{\dot{B}\dot{C}}{BC}\right)B^2.
\end{align}
Solving Eq.(\ref{48}) for acceleration scalar, $a$, using this
solution in second dynamical equation (\ref{43}), we get
\begin{align}\nonumber
&U^*(\mu+P_r) \\\nonumber&=-\left[\left\{
\frac{D_3C'}{A}+\frac{\kappa
C(1+\gamma{R^n})^2(\mu+P_r)}{nR^{n-1}[(1+\gamma{R^n})(2\alpha{R}^n-\beta)-\gamma{R^n}(\alpha
R^n-\beta)]}\right.\right.\\\nonumber
&\times\left(\right.\left.\left.\left.-\frac{2\pi{E^2}}{B^2}-\frac{{\alpha}R^{2n}(1-2n)
-{\beta}R^n(1-n)}{2(1+\gamma{R^n})}+\frac{n\gamma{R}^{2n}(\alpha{R^n}-\beta)}
{2(1+\gamma{R^n})^2}\right.\right.\right.\\\nonumber
&\left.\left.\left. +P_r+\frac{
\psi_{rr}}{A^2}\right)\right\}-\frac{A}{B^2}\right.\left.\times(\mu+P_r)
\left(\frac{R_{0202}}{A^2}-\frac{R_{2323}}{C^2}\right)
\frac{C'}{C}\right]\\\nonumber &
+\left[-q^*+\frac{1}{3}(\sigma_K+\sigma_S-4\Theta)q
\right]\frac{C'}{A}-\frac{C'}{A}\left[P^\dag_r-\frac{1}{A}\left
(\frac{B'}{B}(P_z-P_r)\right.\right.
\\\label{49}
&\left.\left.+\frac{C'}{C}(P_\phi-P_r)\right)+\frac{4\pi{E}}{A^2B^2C}(EC'+E'C)\right],
\end{align}
where $D_3$ represents $f(R)$ extra degrees of freedom. One can
obtain this expression by using Eq.(\ref{44}) in $D_1$. Left hand
side (LHS) is formed through the product of time derivative of
matter four velocity and inertial mass (density). Nevertheless,
right hand side (RHS) of the above equation can be considered as a
combination of three different portions (in three square brackets).
The first square bracket encapsulates effective portion of
relativistic gravitational force (along with inflationary and late
time accelerating $f(R)$ corrections), while the second part
describes parameters controlling dissipative mechanism of the
charged cylindrical collapsing systems. The last term leads to
hydrodynamic force as it is the combination of pressure and
electromagnetic anisotropic gradient entities for evolving
cylindrically symmetric stellar object. It is worthy to stress that
hydrodynamic era can be referred as matter particles that not
necessarily radiate/tranport heat. Equation (\ref{49}) can be
rewritten in Newtonian form as
\begin{align}\nonumber
Acceleration \times Mass~density = Force.
\end{align}
Now we calculate charged cylindrical mass function $m$ framed within
$f(R)$ background. The couple of Reimann tensor components,
$R_{0202}$ and $R_{2323}$ can be written by means of metric $f(R)$
structure functions as
\begin{equation*}\nonumber
\frac{B^2}{3}(2Y_S+Y_T-X_T+X_S+Y_K-X_K)=\frac{1}{A^2}R_{0202}
-\frac{1}{C^2}R_{2323}.
\end{equation*}
Using Eqs.(\ref{23})-(\ref{28}) and feeding back the values of
scalar variables from Eqs.(\ref{A3})-(\ref{A16}), we can modify
above equation as
\begin{align}\nonumber
&\frac{1}{A^2}R_{0202}-\frac{1}{C^2}R_{2323}=\frac{B^2}{3}\left(\mathbb{E}_S
-2\mathbb{E}_K\right)+\frac{{\kappa}B^2}{3f_R}\left(2P_r-P_z\right.\\\nonumber
&
-\frac{6\pi}{B^2}E^2\left.+\frac{{\alpha}R^{2n}(1-2n)-{\beta}R^n(1-n)}
{(1+\gamma{R^n})}+\frac{n\gamma{R}^{2n}(\alpha{R^n}-\beta)}
{(1+\gamma{R^n})^2}\right.\\\label{50} &
-\frac{\mu}{2}-\frac{\psi_{zz}}{B^2}-\frac{\psi_{tt}}{2A^2}+\frac{P_\phi}{2}\left.+\frac{2\psi_{rr}}{A^2}
+\frac{\psi_{\phi\phi}}{2C^2}\right).
\end{align}
Equations (\ref{49}) and (\ref{50}) simultaneously provide
\begin{align}\nonumber
&\frac{\kappa
C(1+\gamma{R^n})^2}{nR^{n-1}[(1+\gamma{R^n})(2\alpha{R}^n-\beta)-\gamma{R^n}(\alpha
R^n-\beta)]}\\\nonumber&\left(P_r-\frac{2\pi}{A^2}E^2+\frac{\psi_{rr}}
{A^2}\right.+\frac{{\alpha}R^{2n}(1-2n)-{\beta}R^n(1-n)}
{2(1+\gamma{R^n})}\\\nonumber
&\left.+\frac{n\gamma{R}^{2n}(\alpha{R^n}-\beta)}
{2(1+\gamma{R^n})^2}\right)-\frac{C}{B^2}\left(\frac{1}{A^2}R_{0202}\right.\left.-\frac{1}
{C^2}R_{2323}\right)\\\nonumber &=\frac{\kappa
C(1+\gamma{R^n})^2}{nR^{n-1}[(1+\gamma{R^n})(2\alpha{R}^n-\beta)-\gamma{R^n}(\alpha
R^n-\beta)]}\\\nonumber &\left(P_r-\frac{2\pi}{B^2} E^2
+\frac{\psi_{rr}}{A^2}\right.\\\nonumber
&\left.+\frac{(1+\gamma{R^n})^2}{nR^{n-1}[(1+\gamma{R^n})(2\alpha{R}^n-\beta)-\gamma{R^n}(\alpha
R^n-\beta)]} \right)-\frac{C}{3}(\mathbb{E}_S\\\nonumber
&-2\mathbb{E}_K)+\frac{\kappa
C(1+\gamma{R^n})^2}{6nR^{n-1}[(1+\gamma{R^n})(2\alpha{R}^n-\beta)-\gamma{R^n}(\alpha
R^n-\beta)]}\\\nonumber &\times\left(\mu-P_\phi
+\frac{6\pi}{B^2}E^2-P_r+2P_z+\frac{\psi_{tt}}{A^2}
-\frac{\psi_{rr}}{A^2}+\frac{2\psi_{zz}}{B^2}\right.\\\label{51}
&\left.-\frac{\psi_{\phi\phi}}{C^2}+\frac{n\gamma{R}^{2n}(\alpha{R^n}-\beta)}
{2(1+\gamma{R^n})^2}+\frac{{\alpha}R^{2n}(1-2n)-{\beta}R^n(1-n)}
{2(1+\gamma{R^n})}\right).
\end{align}
The above relation help us to calculate possible modification of
Misner-Sharp mass formulation \citep{w25} for cylindrical collapsing
object in the presence of electromagnetic field and $f(R)$ extra
curvature terms. For this purpose, we first calculate Eqs.(47) and
(55) mentioned in \citet{newh1} and \citet{w8b}, respectively in
Maxwell-$f(R)$ gravity. Then, we compare these results with
Eq.(\ref{51}), keeping in mind that regenerative pressure gradient
has the same contribution as in spherical symmetrical stellar
systems. Consequently, we found the following peculiar form of mass
function
\begin{align}\nonumber
m&=\frac{\kappa
C^3(1+\gamma{R^n})^2}{6nR^{n-1}[(1+\gamma{R^n})(2\alpha{R}^n-\beta)-\gamma{R^n}(\alpha
R^n-\beta)]}\\\nonumber &\times \left(\mu+\frac{6\pi}{B^2}E^2
-\frac{\psi_{rr}}{A^2}\right.+\frac{\psi_{tt}}{A^2}-P_\phi
+\frac{2\psi_{zz}}{B^2}\\\nonumber
&\left.-\frac{{\alpha}R^{2n}(1-2n)-{\beta}R^n(1-n)}
{2(1+\gamma{R^n})}+\frac{n\gamma{R}^{2n}(\alpha{R^n}-\beta)}
{2(1+\gamma{R^n})^2}\right.\\\label{52}
&\left.-P_r+2P_z-\frac{\psi_{\phi\phi}}{C^2}\right)-\frac{C^3}{3}(\mathbb{E}_S-2\mathbb{E}_K).
\end{align}
We can obtain from Eqs.(\ref{23})-(\ref{28}), (\ref{30}),
(\ref{A3})-(\ref{A16}) and above equation
\begin{align}\nonumber
\frac{3m}{C^3}&=\frac{(1+\gamma{R^n})^2}{2nR^{n-1}[(1+\gamma{R^n})(2\alpha{R}^n-\beta)-\gamma{R^n}(\alpha
R^n-\beta)]}\\\nonumber&\times\left(\mu+P_\phi+\frac{10\pi}{B^2}E^2
\right.-2P_r+P_z-\frac{2\psi_{rr}}{A^2}\\\nonumber
&\left.-\frac{{\alpha}R^{2n}(1-2n)-{\beta}R^n(1-n)}{(1+\gamma{R^n})}
+\frac{n\gamma{R}^{2n}(\alpha{R^n}-\beta)}
{(1+\gamma{R^n})^2}\right.\\\label{53}
&\left.+\frac{\psi_{tt}}{A^2}+\frac{\psi_{\phi\phi}}{C^2}+\frac{\psi_{zz}}{B^2}\right)-(Y_S-2Y_K),\\\nonumber
\frac{3m}{C^3}&=\frac{(1+\gamma{R^n})^2}{2nR^{n-1}[(1+\gamma{R^n})
(2\alpha{R}^n-\beta)-\gamma{R^n}(\alpha R^n-\beta)]}\\\nonumber
&\times\left(\mu+3P_z+\frac{2\pi}{B^2}E^2-3P_\phi+\frac{
\psi_{tt}}{A^2}-\frac{3\psi_{\phi\phi}} {C^2}
+\frac{2\psi_{zz}}{B^2}\right)\\\label{54}&+(X_S-2X_K).
\end{align}
In the presence of electromagnetic field and several scalars
variables, the cylindrical mass function is obtained after using
Eqs.(\ref{23})-(\ref{28}), (\ref{30}), (\ref{A3})-(\ref{A16}),
(\ref{43}), (\ref{45}) and (\ref{46}) as
\begin{align}\nonumber
&-\left[\frac{\kappa(1+\gamma{R^n})^2}{nR^{n-1}
[(1+\gamma{R^n})(2\alpha{R}^n-\beta)-\gamma{R^n}(\alpha
R^n-\beta)]}\right.\\\nonumber
&\left.\times\left(\mu+\frac{2\pi}{B^2}E^2-\frac{{\alpha}R^{2n}(1-2n)
-{\beta}R^n(1-n)}{2(1+\gamma{R^n})}\right.\right.\\\nonumber
&\left.\left. -\frac{n\gamma{R}^{2n}(\alpha{R^n}-\beta)}
{2(1+\gamma{R^n})^2}+\frac{\psi_{tt}}{A^2}\right)\right]^\dag+(X_S+3Y_S+X_K-3Y_K)^\dag\\\nonumber
&=\frac{(1+\gamma{R^n})^2}{nR^{n-1}[(1+\gamma{R^n})(2\alpha{R}^n-\beta)-\gamma{R^n}(\alpha
R^n-\beta)]}\left(q-\frac{\psi_{tr}}{A^2}\right)
\\\nonumber
&(\sigma_K-\Theta-2\sigma_S)-(Y_S+X_S)\frac{3B'}{AB}+\frac{12\kappa\pi{E}}
{A^2B^2Cf_R}\left\{E'C+C'\right.\\\nonumber
&\times\left.\left(E-\frac{k}{C}\right)\right\}+(Y_K-X_K)
\frac{3C'}{AC}-6H^*+6H(\sigma_S-\Theta)\\\label{55}&-3a(Y_S-Y_K-X_S+X_K).
\end{align}
From Eqs.(\ref{53}) and (\ref{54}), we have
\begin{align}\nonumber
&\frac{6m}{C^3}=\frac{\kappa(1+\gamma{R^n})^2}{nR^{n-1}[(1+\gamma{R^n})(2\alpha{R}^n-\beta)-\gamma{R^n}(\alpha
R^n-\beta)]}\\\nonumber&\times\left(\mu+\frac{2\pi}{B^2}E^2\right.\left.-\frac{{\alpha}R^{2n}(1-2n)
-{\beta}R^n(1-n)}{2(1+\gamma{R^n})}+\frac{\psi_{tt}}{A^2}\right.\\\label{55a}
&\left.-\frac{n\gamma{R}^{2n}(\alpha{R^n}-\beta)}
{2(1+\gamma{R^n})^2}\right)-3Y_S-X_K+3Y_K-X_S.
\end{align}
Applying $\dag$ on both sides of Eq.(\ref{55a}) and taking into
account Eq.(\ref{55}), we obtain
\begin{align*}\nonumber
&\left(\frac{6m}{C^3}\right)^\dag=\\\nonumber&-\frac{4\pi\kappa{E}
(1+\gamma{R^n})^2}{nA^2B^2CR^{n-1}[(1+\gamma{R^n})(2\alpha{R}^n-\beta)-\gamma{R^n}(\alpha
R^n-\beta)]}\\\nonumber
&\times\left\{EC'(1-2\pi)+C\left(E'-\frac{4{\pi}EB'}{B}\right)\right\}\frac{3(Y_S+X_S)}
{AB}B'\\\nonumber
&+6H^*-6H(\sigma_S-\Theta)+\frac{3(X_K-Y_K)}{CA}C'\\\nonumber&+\frac{\kappa(1+\gamma{R^n})^2(2\sigma_S-\sigma_k
+\Theta)}{nR^{n-1}[(1+\gamma{R^n})(2\alpha{R}^n-\beta)-\gamma{R^n}(\alpha
R^n-\beta)]}\\\nonumber &\times\left(q-\frac{\psi_{tr}}{A^2}\right)
+3a(X_K+Y_S-X_S-Y_K),
\end{align*}
that yields
\begin{align}\nonumber
&m=\\\nonumber&\frac{C^3}{2}{\int}\left[-\frac{4\pi\kappa{E}
(1+\gamma{R^n})^2}{nA^2B^2CR^{n-1}[(1+\gamma{R^n})(2\alpha{R}^n-\beta)-\gamma{R^n}(\alpha
R^n-\beta)]}\right.\\\nonumber
&\times\left\{EC'(1\left.-2\pi)+C\left(E'-\frac{4{\pi}EB'}{B}\right)\right\}+\frac{(Y_S+X_S)}{B}B'
\right.\\\nonumber &-2AH(\sigma_S-\Theta)
+2AH^*+\frac{(X_K-Y_K)}{C}C'\\\nonumber&\left.+\frac{\kappa
A(1+\gamma{R^n})^2(2\sigma_S+\Theta-\sigma_K)}{3nR^{n-1}[(1+\gamma{R^n})(2\alpha{R}^n-\beta)-\gamma{R^n}(\alpha
R^n-\beta)]}\left(q\right. \right.\\\label{56} &\left.
\left.-\frac{\psi_{tr}}{A^2}\right)+aA(Y_S-X_S+X_K-Y_K)\right]dr+\frac{C^3\zeta(t)}{6},
\end{align}
in which $\zeta$ is an integration function.

One can easily observe from Eq.(\ref{46}) that if our system, that
is coupled with non-radiating isotropic matter configurations, is
conformally flat, then the homogeneity in the energy density is
controlled by $f(R)$ degrees of freedom. Further, if one takes
current cosmological value of curvature tensor, then system will
encapsulate regular energy density throughout over the celestial
object with conformally flat environment. At that case, the last
portion of the above equation makes the mass function for the
regular configurations of celestial cylindrical object. The present
version of mass function (mentioned at Eq.(\ref{57})) points out
that how much system scalar variables and electric charge are
important in its formulation. Substituting $X_S,~Y_K,~Y_S,~X_K$ in
Eq.(\ref{56}), we get
\begin{align}\nonumber
m&=\frac{C^3}{2}{\int}\left[\frac{-\kappa(1+\gamma{R^n})^2}{nR^{n-1}
[(1+\gamma{R^n})(2\alpha{R}^n-\beta)-\gamma{R^n}(\alpha
R^n-\beta)]}\right.\\\nonumber
&\times\left(P_z-P_r+\frac{4\pi}{B^2}\left.E^2+
\frac{\psi_{zz}}{B^2}-\frac{\psi_{rr}}
{A^2}\right)\frac{B'}{B}-\left\{EC'(1-2\pi)\right.\right.\\\nonumber
&+C\left.\left(E'-\frac{4{\pi}EB'}{B}\right)\right\}\\\nonumber
&\times\frac{4\pi\kappa{E}
(1+\gamma{R^n})^2}{nA^2B^2CR^{n-1}[(1+\gamma{R^n})(2\alpha{R}^n-\beta)-\gamma{R^n}(\alpha
R^n-\beta)]}\\\nonumber &-2[H(\sigma_S-\Theta)-H^*]A
-2\mathbb{E}_K\frac{C'}{C}\\\nonumber&+\frac{\kappa
A(1+\gamma{R^n})^2}{3nR^{n-1}[(1+\gamma{R^n})(2\alpha{R}^n-\beta)-\gamma{R^n}(\alpha
R^n-\beta)]}\\\label{57}
&\left.\times\left(q+\frac{\psi_{tr}}{A^2}\right)(2\sigma_S+\Theta-\sigma_K)\right]dr+\frac{C^3\zeta(t)}{6}.
\end{align}
in which influence of tidal forces, dark source extra curvature
terms, radiating parameters, local anisotropic pressure, shear,
expansion and modified version of scalar functions can easily be
exhibited. It can be observed that for conformally flat isotropic
non-dissipative cylinder, the inhomogeneity in the energy density is
disturbed by electric charge and realistic $f(R)$ corrections. Then
last term of the above equation holds fundamental importance in the
measurement of mass function of the uncharged perfect cylindrical
compact object with $f(R)$ corrections.

The problem of collapsing systems can be well discussed by joining
two appropriate geometries of interior and exterior spacetimes. For
this purpose, we consider that system relativistic motion is
characterized by three dimensional timelike surface represented by
$\Omega$. Thus, $\Omega$ demarcated our manifold into interior and
exterior portions, denoted, respectively, by $\mathcal{V}^-$ and
$\mathcal{V}^+$. The spacetime for $\mathcal{V}^+$ is \cite{chao1}
\begin{equation}\label{2n}
ds^2_+=\left(-\frac{2M}{r}\right)d\nu^2
+2d{\nu}d{r}-r^2(d\phi^2+\zeta^2dz^2),
\end{equation}
where $M$ is a cylindrical gravitating mass and $\zeta$ indicates
arbitrary constant. The spacetime for $\mathcal{V}^-$ is given in
Eq.(\ref{5}). For continuous matching of $\mathcal{V}^-$ and
$\mathcal{V}^+$, we use \citet{dar1} and \citet{seno1} matching
criteria. Darmois conditions require continuity of both line
elements and extrinsic curvature over $\Omega$ (see for details
\citet{chan1}), while Senovilla matching conditions requires
\begin{align}\label{24n}
R|_-^+&=0,\quad f_{,RR}[\partial_\nu R|_-^+=0,\quad f_{,RR}\neq0.
\end{align}
These couple of constraints assert that Ricci invariant must be
continuous over $\Omega$ even for matter thin shells.
\begin{figure} \centering
\epsfig{file=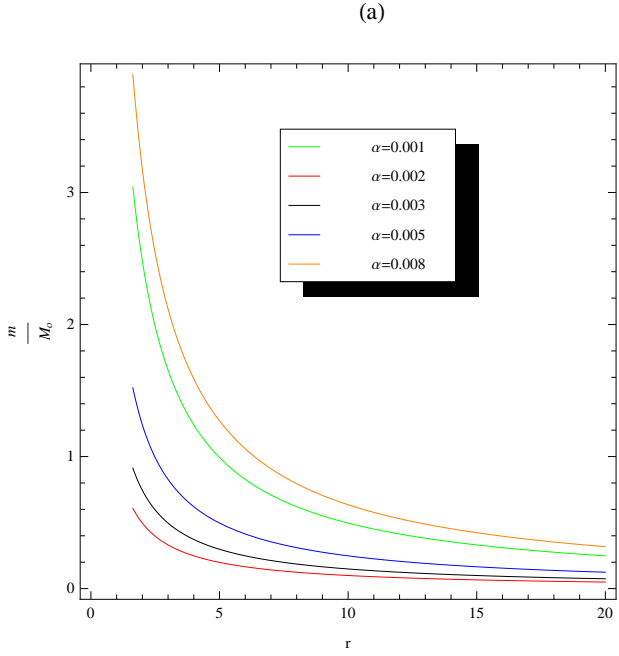,width=.98\linewidth}
 \caption{Mass radius relationship diagram for relativistic cylinders in $f(R)$ gravity, with
 different values of $\alpha$ alongwith $n=1$, $\beta=0=\gamma$.}
\end{figure}
\begin{figure} \centering
\epsfig{file=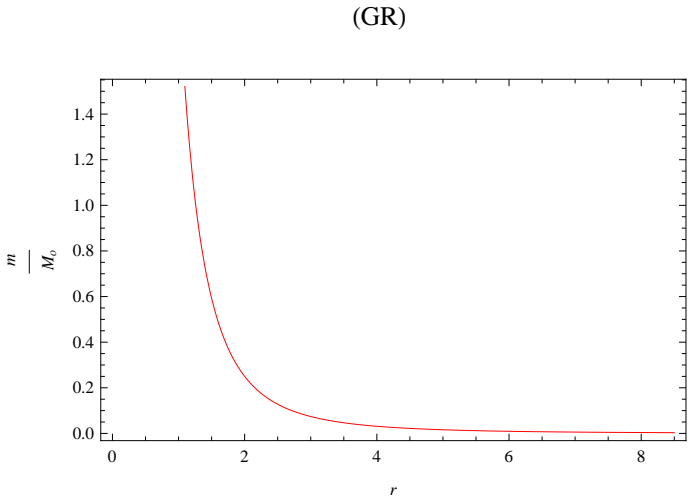,width=1.09\linewidth}
 \caption{Mass radius relationship diagram in GR.}
\end{figure}
\begin{figure} \centering
\epsfig{file=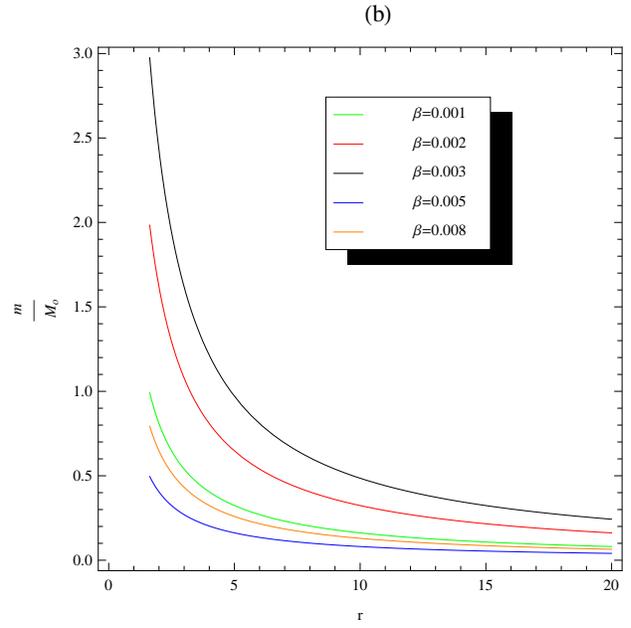,width=.98\linewidth}
 \caption{Mass radius relationship diagram for relativistic cylinders in $f(R)$ model, with
 different values of $\beta$ alongwith $n=1$, $\gamma=0$.}
\end{figure}
\begin{figure} \centering
\epsfig{file=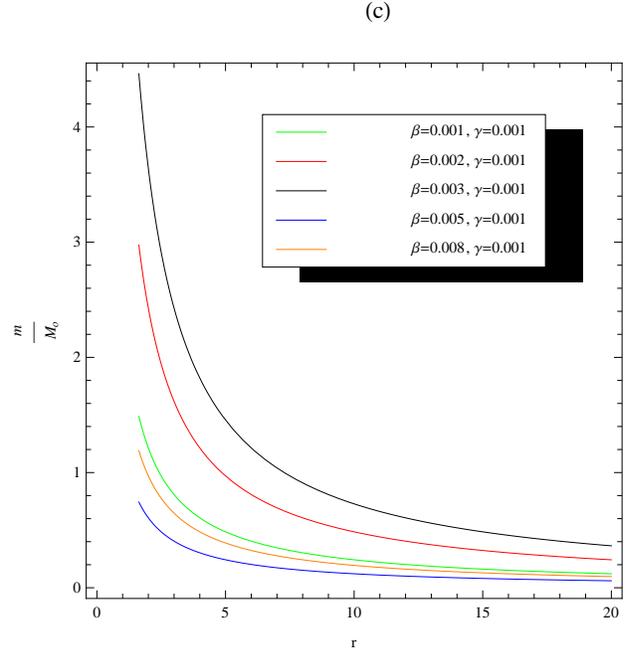,width=.98\linewidth}
 \caption{Mass radius relationship diagram for relativistic cylinders in $f(R)$ model, with
 different values of $\beta$ and  $\gamma$ alongwith $n=1$.}
\end{figure}

\subsection[]{Gravity Induced by Specific Versions of $f(R)$
Models} Using \citet{seno1} and \citet{dar1} criteria over the
hypersurface in Eq.(\ref{57}), we found some relationships between
mass and radius for different values of parameters mentioned in
(\ref{44}) $f(R)$ model. These parameters allow us to set
up various cylindrical compact models controlled by some specific $f(R)$ configurations.\\\\
\textbf{Model 1.} First, we consider the simple form of
Eq.(\ref{44}) given by
\begin{align}\label{11ab}
f(R)={\alpha}R^{2},
\end{align}
where $\alpha$ is positive This type of model was claimed to
describe early time cosmic inflation due to an additional $\alpha
R^2$ term. These corrections in the field equations \citep{ya33b}
are consistent with temperature anisotropies noticed in cosmic
microwave background. Now, we shall see the effects of $f(R)$
corrections in the dynamical behavior of cylindrical anisotropic
matter distribution. We found mass radius relationships with
different values of $\alpha$ and $\beta=\gamma=0$ mentioned in
figure \textbf{1}. It is seen that the maximum $\frac{m}{M_\odot}$
ratio of relativistic cylinder is about $3.4$ against 1.1 value of
radius. In GR, we also found mass radius relationship and concluded
that the maximum $\frac{m}{M_\odot}$ ratio is 1.4 for 1.1 value of
radius as seen in figure \textbf{2}. Thus, such modified models
yield more compact relativistic cylinders with smaller radii than
GR. Our this result supports the analysis of
\cite{asta1}.\\\\
\textbf{Model 2.} It would be interesting to analyze $f(R)$
corrections of the following type
\begin{align}\label{12ab}
f(R)={\alpha}R^{2}-\beta R,
\end{align}
where $\alpha$ and $\beta$ are positive. The models of this type
could allow to deal with some unexplored cylindrical
self-gravitating dissipative systems that was escaped from standard
gravitational theory. We found mass radius relationship by
considering above mentioned corrections in the action. We conclude
that for different values of $\beta$ with fixed value of $\alpha$
and $\gamma=0$, the maximum $\frac{m}{M_\odot}$ ratio of cylindrical
stellar objects is $3.0$ for about 1.6 units of radius as shown in
figure \textbf{3}. This indicates that very massive stellar compact
objects that could not be obtained as solutions of the standard
stellar structures theory could be achieved in the realm of above mentioned $f(R)$
gravity.\\\\
\textbf{Model 3.} Now, we consider the following specific form of
$f(R)$ correction in the action
\begin{align}\label{13ab}
f(R)=\frac{{\alpha}R^{2}-\beta R}{1+\gamma R},
\end{align}
where $\alpha,~\beta$ and $\gamma$ are positive constants. Here, we
found mass radius relation diagram (figure \textbf{4}) for different
values of $\beta$ and $\gamma$ with fixed value of $\alpha$. We
conclude maximum $\frac{m}{M_\odot}$ ratio to be $4.5$ against about
$1.5$ units of radius. This $\frac{m}{M_\odot}$ ratio is found to be
greater than that obtained in the above couple of $f(R)$ models.
Further, it can be analyzed from figures \textbf{2-4} that wide
range of massive compact objects occur in $f(R)$ gravity. This fact
can be interpreted as there is a possibility of the occurrence of
more massive stable cylindrical objects in modified gravity (with
some specific $f(R)$ models). These results support the consequences
of \citet{fari1}.

\section[]{$f(R)$ Dynamical-Transport Equation}

The transport expression from casual radiating theory is
\begin{align}\nonumber
\tau{q}^*&=-\frac{1}{2}\eta{q}\Omega^2\left(\frac{\tau}{{\eta}\Omega^2}\right)^*
-\eta(\Omega^\dag+\Omega a)-q-\frac{1}{2}{\tau}q\Theta,
\end{align}
where $\tau$ indicates relaxation time, $\eta$ is a thermal
conductivity, while $\Omega$ stands for temperature. On setting
$\tau=0$ in the above equation, one can get Eckart-Landau equation
\citep{b30a, b30b}. The non-vanishing independent component is
\begin{equation}\label{58}
{\tau}q^*+q=-\eta(\Omega^\dag+\Omega a)-\frac{1}{2}\eta
\Omega^2q\left(\frac{\tau}{\eta\Omega^2}\right)^*
-\frac{1}{2}q\tau\Theta.
\end{equation}
After using Eqs.(\ref{51}), (\ref{52}) and (\ref{58}) in (\ref{49}),
we get
\begin{align}\nonumber
&(\mu+P_r)\left\{1-\frac{{\eta}\Omega}{\tau(\mu+P_r)}\right\}U^*
=-\left\{\left(P_r+\frac{{\alpha}R^{2n}(1-2n)}{2(1+\gamma{R^n})}\right.\right.\\\nonumber
&-\left.\left.\frac{{\beta}R^n(1-n)}{2(1+\gamma{R^n})}+\frac{n\gamma{R}^{2n}(\alpha{R^n}-\beta)}
{2(1+\gamma{R^n})^2}-\frac{2\pi}{A^2}E^2+\frac{
\overset{~~~(D)}{T_{11}}}{A^2}\right)\right.\\\nonumber
&\left.\times[\kappa C^3(1+\gamma{R^n})^2][2nR^{n-1}[(1+\gamma{R^n})
(2\alpha{R}^n-\beta)-\gamma{R^n}\right.\\\nonumber
&\left.\times(\alpha
R^n-\beta)]]^{-1}+m\right\}\frac{(\mu+P_r)}{C^2}\left\{1-\frac{{\eta}
\Omega}{\tau(\mu+P_r)}\right\} +\frac{4{\pi}E}{A^3B^2C}\\\nonumber
&\times(CE'+EC')C' +\frac{C'}{A}\left[-P^\dag_r
+\frac{1}{A}\left\{(P_\phi-P_r)\frac{C'}{C}+(P_z\right.\right.\\\nonumber
&\left.\left.-P_r)\frac{B'}{B}\right\}
\right]+\frac{C'}{A}\left[q\left\{\frac{1}{\tau}+\frac{{\eta}\Omega^2}{2\tau}\left(\frac{\tau}
{{\eta}\Omega^2}\right)^*+\frac{(\sigma_S+\sigma_K)}{3}\right.\right.\\\nonumber
&\left.\left.-\frac{5}{6}\Theta\right\}
+\frac{{\eta}\Omega^\dag}{\tau}\right]-\frac{D_3C'}{A},
\end{align}
which can be recast as
\begin{align}\nonumber
&(\mu+P_r)\left(1-\mathfrak{a}\right)U^*
=F_{grav}(1-\mathfrak{a})+\frac{C'}{A}\left[-P^\dag_r
+\frac{1}{A}\left\{(P_\phi\right.\right.\\\nonumber
&-\left.\left.P_r)\frac{C'}{C}+(P_z-P_r)\frac{B'}{B}\right\}-D_1
\right]+\frac{C'}{A}\left[q\left\{\frac{1}{\tau}+\frac{{\eta}\Omega^2}{2\tau}\right.\right.\\\nonumber
&\times\left.\left.\left(\frac{\tau}
{{\eta}\Omega^2}\right)^*+\frac{(\sigma_S+\sigma_K)}{3}
-\frac{5}{6}\Theta\right\}+\frac{{\eta}\Omega^\dag}{\tau}\right]
+\frac{4{\pi}E}{A^3B^2C}\\\label{59} &\times(CE'+EC')C',
\end{align}
where
\begin{align}\nonumber
&F_{grav}=-\frac{(\mu+P_r)}{C^2}\left\{\left(P_r+\frac{{\alpha}R^{2n}(1-2n)
-{\beta}R^n(1-n)}{2(1+\gamma{R^n})}\right.\right.\\\nonumber
&+\left.\left.\frac{n\gamma{R}^{2n}(\alpha{R^n}-\beta)}
{2(1+\gamma{R^n})^2}-\frac{2\pi}{A^2}E^2+\frac{
\overset{~~~(D)}{T_{11}}}{A^2}\right)\right.\\\nonumber
&\times\left.\frac{[\kappa
C^3(1+\gamma{R^n})^2]}{[2nR^{n-1}[(1+\gamma{R^n})(2\alpha{R}^n-\beta)-\gamma{R^n}(\alpha
R^n-\beta)]]}+m\right\},\\\nonumber
&\mathfrak{a}=\frac{{\eta}\Omega}{\tau(\mu+P_r)}.
\end{align}
This equation shows that thermoinertial effects tends to reduce the
effective inertial quantity of matter distribution of cylindrical
stellar object with metric extra curvature terms.

\section[]{Static Charged Anisotropic/Isotropic Cylinders}

Here, we investigate the role of extended versions of structure
scalars in the modeling of static non-rotating cylindrical systems
in the presence of electromagnetic field and metric $f(R)$
corrections. In this section, we shall write the notation $\chi_is$
for functions/contants arising after integration. The metric $f(R)$
field equations for static cylindrically symmetric spacetime modeled
with anisotropic matter distribution are
\begin{align}\nonumber
&\alpha_2-\frac{B'^2}{B^2}-\frac{C''}{C}-\frac{B''}{B}=\frac{\kappa}{f_R}\left[\mu{A^2}
+\frac{2\pi{A^2}}{B^2}E^2+\frac{A^2}{\kappa}\right.\\\label{60}
&\times\left.\left\{\frac{f''_R}{A^2} -\frac{A'}{A}
\frac{f'_R}{A^2}+\gamma_2\right\}\right],\\\label{61}
&\frac{B'^2}{B^2}+\alpha_2
=\frac{\kappa}{f_R}\left[P_r{A^2}-\frac{2\pi{A^2}}{B^2}E^2+\frac{A^2}{\kappa}\left\{-\gamma_2-
\frac{A'}{A}\frac{f'_R}{A^2}\right\}\right],\\\label{62}
&\frac{C''}{C}\frac{B^2}{A^2}+\beta_2B^2=\frac{\kappa}{f_R}\left[P_z{B^2}+{2\pi}E^2+\frac{B^2}{\kappa}\left\{
\delta_2-\frac{C'f'_R}{CA^2}\right\}\right],\\\label{63}
&\frac{B''}{B}\frac{C^2}
{A^2}+\beta_2C^2=\frac{\kappa}{f_R}\left[P_\phi{C^2}+\frac{2\pi{C^2}}{B^2}E^2+\frac{C^2}{\kappa}\left\{
\delta_2-\frac{B'f'_R}{BA^2}\right\} \right],
\end{align}
where
\begin{align}\nonumber
&\alpha_2=\frac{A'}{A}\left(\frac{B'}{B}+\frac{C'}{C}\right),
\quad\gamma_2=\frac{Rf_R-f}{2}+\left(\frac{B'}{B}+\frac{C'}{C}\right)
\frac{f'_R}{A^2},\\\nonumber
&\beta_2=\left(\frac{A''}{A}-\frac{A'^2}{A^2}\right)\frac{1}{A^2},\quad
\delta_2=\frac{f-Rf_R}{2}-\frac{f''_R}{A^2}.
\end{align}
Now we would take some auxiliary terms,
$\psi_1\equiv\frac{A'}{A},~\psi_2\equiv\frac{B'}{B},~\psi_3\equiv\frac{C'}{C},~
\psi_4\equiv\frac{f'_R}{f_R}$. Under this formulation,
Eqs.(\ref{61})-(\ref{63}) with viable $f(R)$ model (\ref{44}) become
\begin{align}\nonumber
&\frac{A^2\kappa(1+\gamma{R^n})^2}{nR^{n-1}[(1+\gamma{R^n})(2\alpha{R}^n-\beta)-\gamma{R^n}(\alpha
R^n-\beta)]}\left(\mu-\frac{n\gamma{R}^{2n}} {2}\right.\\\nonumber
&\times\left. \frac{(\alpha{R^n}-\beta)}
{2(1+\gamma{R^n})^2}+\frac{{\alpha}R^{2n}(1-2n)-{\beta}R^n(1-n)}{2(1+\gamma{R^n})}
+\frac{\phi_{tt}}{A^2}\right.\\\nonumber
&\left.+\frac{2{\pi}E^2}{B^2}\right)=-\psi_2'-\psi_2^2-\psi_3'-\psi_3^2
+\psi_1\psi_2+\psi_1\psi_3\\\label{63} &-\psi_2 \psi_3,\\\nonumber
&\frac{A^2\kappa(1+\gamma{R^n})^2}{nR^{n-1}[(1+\gamma{R^n})(2\alpha{R}^n-\beta)-\gamma{R^n}(\alpha
R^n-\beta)]}\left(P_r\right.\\\nonumber
&\left.+\frac{n\gamma{R}^{2n}(\alpha{R^n}-\beta)}
{2(1+\gamma{R^n})^2}-\frac{{\alpha}R^{2n}(1-2n)-{\beta}R^n(1-n)}{2(1+\gamma{R^n})}\right.\\\label{64}
&\left. +\frac{\phi_{rr}}{A^2}-\frac{2{\pi}E^2}{B^2}\right)=
\psi_1\psi_2+\psi_1\psi_3+\psi_2\psi_3,\\\nonumber
&\frac{A^2\kappa(1+\gamma{R^n})^2}{nR^{n-1}[(1+\gamma{R^n})(2\alpha{R}^n-\beta)-\gamma{R^n}(\alpha
R^n-\beta)]}\left(P_z\right.\\\nonumber
&\left.+\frac{n\gamma{R}^{2n}(\alpha{R^n}-\beta)}
{2(1+\gamma{R^n})^2}-\frac{{\alpha}R^{2n}(1-2n)-{\beta}R^n(1-n)}{2(1+\gamma{R^n})}\right.\\\label{65}
&\left. +\frac{\phi_{zz}}{A^2}+\frac{2{\pi}E^2}{B^2}\right)=
\psi_1'+\psi_3'+\psi_3^2,\\\nonumber
&\frac{A^2\kappa(1+\gamma{R^n})^2}{nR^{n-1}[(1+\gamma{R^n})(2\alpha{R}^n-\beta)-\gamma{R^n}(\alpha
R^n-\beta)]}\left(P_\phi\right.\\\nonumber
&\left.+\frac{n\gamma{R}^{2n}(\alpha{R^n}-\beta)}
{2(1+\gamma{R^n})^2}-\frac{{\alpha}R^{2n}(1-2n)-{\beta}R^n(1-n)}{2(1+\gamma{R^n})}\right.\\\label{66}
&\left. +\frac{\phi_{zz}}{A^2}+\frac{2{\pi}E^2}{B^2}\right)=
\psi_1'+\psi_2' +\psi_2^2,
\end{align}
where $\phi_{\alpha\beta}$ are dark source terms corresponding to
static cylindrical system and can be evaluated from the
corresponding values of $\psi_{\alpha\beta}$. It can be observed
that all $\psi_{\alpha\beta}$ reduce to their respective
$\phi_{\alpha\beta}$ under static environment of the metric
variables. The couple of scalar variables corresponding to electric
part of the Weyl tensor under auxiliary variables turn out to be
\begin{align}\label{67}
&\mathbb{E}_S=\frac{1}{2A^2}\left(-\psi_1'+\psi_3'+\psi_3^2+\psi_1\psi_2
-\psi_2\psi_3-\psi_1\psi_3\right),\\\label{68}
&\mathbb{E}_K=\frac{1}{2A^2}\left(-\psi_1'+\psi_2'+\psi_2^2-\psi_1\psi_2
+\psi_1\psi_3-\psi_2\psi_3\right).
\end{align}

The structure scalar $Y_T$ can be reshaped by means of auxiliary
variables as
\begin{equation}
\psi_1'+\psi_1\psi_2+\psi_1\psi_3=Y_TA^2.\\\label{69}
\end{equation}
Equations (\ref{63})-(\ref{66}) give
\begin{align}\nonumber
&\frac{\kappa{A^2}(1+\gamma{R^n})^2}{nR^{n-1}[(1+\gamma{R^n})(2\alpha{R}^n-\beta)-\gamma{R^n}(\alpha
R^n-\beta)]}\left(P_\phi+\frac{4\pi{E^2}}{B^2}\right.\\\label{70}
&\left.-P_r+\frac{\phi_{\phi\phi}}{C^2}-\frac{\phi_{rr}}
{A^2}\right)=\psi_1'+\psi_2'+\psi_2^2-\psi_1\psi_2-\psi_1\psi_3-\psi_2\psi_3
,\\\nonumber
&\frac{\kappa{A^2}(1+\gamma{R^n})^2}{nR^{n-1}[(1+\gamma{R^n})(2\alpha{R}^n-\beta)-\gamma{R^n}(\alpha
R^n-\beta)]}\left(P_z+\frac{4\pi{E^2}}{B^2}\right.\\\label{71}
&\left.-P_r+\frac{\phi_{zz}}{B^2}-\frac{\phi_{rr}}
{A^2}\right)=\psi_1'+\psi_3'+\psi_3^2-\psi_1\psi_2-\psi_1\psi_3-\psi_2\psi_3
,\\\nonumber
&\frac{\kappa{A^2}(1+\gamma{R^n})^2}{nR^{n-1}[(1+\gamma{R^n})(2\alpha{R}^n-\beta)-\gamma{R^n}(\alpha
R^n-\beta)]}\left(P_\phi -P_z\right.\\\label{72}
&\left.+\frac{\phi_{\phi\phi}}{C^2}-\frac{
\overset{~~~(D)}{\mathcal{T}_{22}}}{B^2}\right)=\psi_2'+\psi_2^2-\psi_3'+\psi_3^2.
\end{align}
Using Eqs.(\ref{24}), (\ref{25}), (\ref{67}), (\ref{68}), (\ref{70})
and (\ref{71}), we obtain
\begin{align}\label{73}
\psi_1'-\psi_1\psi_2=-Y_SA^2,\quad \psi_1'+\psi_1\psi_3=-Y_KA^2,
\end{align}
whose integrations, respectively, give
\begin{align*}
A={\chi_1}e^{{\int}B\left(\int\frac{-Y_SA^2}{B}dr\right)dr},\quad
A={\chi_2}e^{{\int}C\left(\int\frac{-Y_KA^2}{C}dr\right)dr},
\end{align*}
which gives $B=B(A)$ and $C=C(A)$ or
$\psi_1=\psi_1(\psi_2),~\forall~Y_S$ and
$\psi_1=\psi_1(\psi_3),~\forall~Y_K$. This asserts that any
auxiliary variable can be written in terms of the other one. For
instance, RHS of Eq.(\ref{72}) can be reexpressed containing the
derivatives of $\psi_1$. Once $\psi_1$ is calculated (with specific
given value of $f(R)$ model) then differences of pressure gradients
can easily be evaluated by using Eqs.(\ref{70}) and (\ref{71}). One
can reduce the complexity of this differential equation by taking
current value of Ricci invariant in $f(R)$ model. Further,
Eqs.(\ref{25}), (\ref{27}) and (\ref{44}) provide
\begin{align}\nonumber
&\frac{\kappa(1+\gamma{R^n})^2}{nR^{n-1}[(1+\gamma{R^n})(2\alpha{R}^n-\beta)-\gamma{R^n}(\alpha
R^n-\beta)]}\left(P_\phi-P_r\right.\\\label{74}
&\left.+\frac{4\pi{E^2}}{B^2}+\frac{\phi_{\phi\phi}}{C^2}-\frac{\phi_{rr}}{A^2}\right)
=-(X_K+Y_K),
\end{align}
while Eqs.(\ref{24}), (\ref{26}) and (\ref{44}) give
\begin{align}\nonumber
&\frac{\kappa(1+\gamma{R^n})^2}{nR^{n-1}[(1+\gamma{R^n})(2\alpha{R}^n-\beta)-\gamma{R^n}(\alpha
R^n-\beta)]}\left(P_z-P_r\right.\\\label{75}
&\left.+\frac{4\pi{E^2}}{B^2}+\frac{\phi_{zz}}{B^2}-\frac{\phi_{rr}}{A^2}\right)
=-(X_S+Y_S).
\end{align}
Thus, it is concluded that that any static cylindrical stellar
solution coupled with anisotropic standard and metric $f(R)$
effective fluid distributions can be obtained from the set of any of
two structure scalar triplets, i.e., $(Y_K,~Y_S,~X_S)$ or
$(Y_K,~Y_S,~X_K)$.

Next, we reduce the complexity of the system and consider static
cylindrical celestial object filled with ideal matter configurations
along with $f(R)$ corrections. Then, field equations
(\ref{60})-(\ref{62}) provide
\begin{align}\nonumber
&\alpha_2-\frac{B'^2}{B^2}-\frac{C''}{C}-\frac{B''}{B}=\frac{\kappa}{f_R}\left[\mu{A^2}
+\frac{2\pi{A^2}}{B^2}E^2+\frac{A^2}{\kappa}\right.\\\label{76}
&\times\left.\left\{\frac{f''_R}{A^2} -\frac{A'}{A}
\frac{f'_R}{A^2}+\gamma_2\right\}\right],\\\label{77}
&\frac{B'^2}{B^2}+\alpha_2
=\frac{\kappa}{f_R}\left[P{A^2}-\frac{2\pi{A^2}}{B^2}E^2+\frac{A^2}{\kappa}\left\{-\gamma_2-
\frac{A'}{A}\frac{f'_R}{A^2}\right\}\right],\\\label{78}
&\frac{C''}{C}\frac{B^2}{A^2}+\beta_2B^2=\frac{\kappa}{f_R}
\left[P{B^2}+{2\pi}E^2+\frac{B^2}{\kappa}\left\{
\delta_2-\frac{C'f'_R}{CA^2}\right\}\right],\\\label{79}
&\frac{B''}{B}\frac{C^2}
{A^2}+\beta_2C^2=\frac{\kappa}{f_R}\left[P{C^2}+\frac{2\pi{C^2}}{B^2}E^2+\frac{C^2}{\kappa}\left\{
\delta_2-\frac{B'f'_R}{BA^2}\right\} \right].
\end{align}
Substituting the value of $P$ from Eq.(\ref{78}) in Eq.(\ref{79}),
we get
\begin{align}\label{80}
\frac{C''}{C}+\frac{C'f'_R}{Cf_R}=\frac{B''}{B}+\frac{B'f'_R}{Bf_R},
\end{align}
which by means of auxiliary functions can be recasted as
\begin{align}\label{81}
{\psi_3'}+\psi_3^2+\psi_3\psi_4=\psi_2'+\psi_2^2+\psi_2\psi_4.
\end{align}
This is the Ricatti equation whose generic solution can be written
as
\begin{align}\label{82}
\psi_3=\psi_2+\frac{1}{k(r)},
\end{align}
where
\begin{align}\label{83}
k(r)=e^{{\int(2\psi_2+\psi_4)dr}}\left[{\int}e^{{\int}-(2\psi_2
+\psi_4)dr}dr+\chi_3\right].
\end{align}
Using $k$ from above equation in Eq.(\ref{82}) and after some
manipulation, we obtain
\begin{align}\label{84}
&C=B{\gamma}\exp\left[B^2f_R\left(\chi_3\right.\right.\\\nonumber
&\left.\left.+{\int}\frac{B^2(1+\gamma{R^n})^2}
{nR^{n-1}[(1+\gamma{R^n})(2\alpha{R}^n-\beta)-\gamma{R^n}(\alpha
R^n-\beta)]}{dr}\right)\right],
\end{align}
in which one should need to consider regularity conditions, i.e.,
$C(t,0)=0$, for feasible model evaluation.

\section[]{Conclusion}

The $f(R)$ gravity models, in which the Lagrangian for the
gravitation is a function of curvature scalar, has attracted the
attention of many researchers. However, to find the solutions of
such field equations is still rare. In this manuscript, we have
presented a procedure indicating how one can write the field
equations using some set of scalar function in static form. The
philosophy of this work is to demonstrate dynamical variables for
cylindrical collapsing object and to present a procedure to find the
solutions of field equations with the properties of modified gravity
having positive and negative powers of curvature. We consider that
the cylindrical body is filled with anisotropic matter in the
presence of dissipation with diffusion approximation under the
influence of electromagnetic field. The modified field equations and
conservation laws are explored for the systematic construction of
our work. We explore the expressions for the shear tensor as well as
the Weyl tensor and analyze some crucial aspects as:
\begin{itemize}
\item  There are two scalars associated with the shear tensor unlike the
spherical case where it has only one such scalar. For the shearfree
system these two scalars must disappear.
\item The Weyl tensor can be decomposed into two constituent tensors namely, its
electric and magnetic parts. It is observed that magnetic part
vanishes in the spherical fluid configuration due to symmetry of the
problem, but in cylindrical star formalism, the magnetic part is
non-zero implying a crucial role in our dynamical analysis.
\end{itemize}

Next, we have disintegrated the Riemann tensor using comoving
vectors which are orthogonal to each other. We found three tensors
called the electric, magnetic and dual of the Riemann tensor.
Further, these tensors are written in terms of its trace and
trace-free scalar parts called the structure scalars. Using the
modified field equations, we have obtained these scalar parts in
terms of material variables like anisotropic stresses, energy
density, heat flux and particularly electric charge and $f(R)$
higher curvature terms. It is found that there are three scalar
functions associated with the electric and dual part of the Riemann
tensor while two scalars are associated with its magnetic part. We
would like to stress here that magnetic part of the Riemann tensor
is due to the presence of magnetic part of Weyl tensor as well as
presence of heat radiations and if the magnetic part of the Weyl
tensor vanishes in dissipation less case then it leads to the
vanishing of Riemann tensor's magnetic part. It is interesting to
note that in the spherical case both the magnetic parts of the Weyl
as well as the Riemann tensors are zero for non-dissipative systems.
A particular viable $f(R)$ model is used to examine the role of dark
source terms on the evolution of these structure scalars. We have
also explored two evolution equations using these structure scalars
to investigate their role on the homogeneity of cylindrical object.

We have obtained an expression for the mass function in $f(R)$
gravity model for cylindrical object in comparison with the
Misner-Sharp mass in spherical system. We have also explained this
mass using the modified structure scalars obtained in the previous
section. After that we found mass radius relationship plots for
different parameter values. We found that one can get even more
compact and massive cylindrical stellar systems (as compared to GR)
for modified gravities induced by $f(R)=\alpha R^2,~f(R)=\alpha
R^2-\beta R$ and $f(R)=\frac{\alpha R^2-\beta R}{1+\gamma R}$
models.

Since our system is dealing with the heat dissipation in diffusion
approximation so we have explored the transportation of heat
phenomena through second order thermodynamical theory defined by
M\"{u}ller and Israel. We have coupled the transport equation with
the dynamical equation obtained from the second non-zero equation of
the conservation law to investigate the collapsing behavior of the
system as given in Eq.(\ref{59}). We found $\mathfrak{a}$ to be
responsible for the decrease and increase in the gravitational mass
of the system and indicate how thermal effects reduce the effective
inertial mass. We observe three main possibilities on this value for
the collapsing behavior as follows:
\begin{itemize}
\item If $\mathfrak{a}\rightarrow 1$, then the inertial mass of
the system approaches to zero value indicating null system inertial
force thereby leading the system towards gravitational attraction
and consequently causes the collapse.
\item If $\mathfrak{a}>1$, then it makes increment in the
inertial mass of the system. This behavior is consistent with the
equivalence principle indicating the expanding behavior of the
system.
\item When $0<\mathfrak{a}<1$, then inertial mass density keeps on moving in
the decreasing state.
\end{itemize}

Since to find the solutions of the modified field equations is huge
task and particularly with dissipation and electromagnetic effects
so in the last section we have presented a framework to explore the
solutions of $f(R)$ field equations in the static background.

It is worthy to stress that all GR already established results
\citep{ya31} can be obtained by taking $f(R)\rightarrow R$ and
$s=0$.

\vspace{0.25cm}

{\bf Acknowledgments}

\vspace{0.25cm}

The authors would like to thank the anonymous reviewer for valuable
and constructive comments and suggestions to improve the quality of
the paper. This work was supported by University of the Punjab,
Lahore-Pakistan through research project in the fiscal year
2015-2016 (Z.Y.).

\appendix

\section[]{}

The $f(R)$ corrections in Eqs.(\ref{41}) and (\ref{42}) are
\begin{align}\nonumber
&D_0=\frac{1}{\kappa}\left[\left\{\left(\dot{f_R}\frac{A'}{A}
+f'_R\frac{\dot{A}}{A}-\dot{f'_R}\right)\frac{1}{A^4}\right\}_{,1}
+\frac{1}{A^2}\left\{\frac{f'_R}{A^2} \right.\right.\\\nonumber
&\times\left.\left.\left(\frac{B'}{B}-\frac{A'}{A}+\frac{C'}{C}\right)
-\left(\frac{\dot{C}}{C}+\frac{\dot{A}}{A}+\frac{\dot{B}}{B}\right)
\frac{\dot{f_R}}{A^2}+\frac{f''_R}{A^2}\right.\right.\\\nonumber
&\left.\left.+\left(\frac{Rf_R-f}{2}\right)\right\}_{,0}+\frac{\dot{A}}{A}
\left\{\frac{\ddot{f_R}}{A^2}+\frac{f''_R}{A^2}-2\frac{A'f'_R}{A^3}-2\frac{\dot{A}
\dot{f_R}}{A^3}\right\}\right.\\\nonumber
&\left.\times\frac{1}{A^2}+\frac{\dot{B}}{B}\left\{\frac{
\ddot{f_R}}{A^2}-\left(\frac{\dot{B}}{B}+\frac{\dot{A}}{A}\right)
\frac{\dot{f_R}}{A^2}+\frac{f'_R}{B^2}
\left(\frac{B'}{B}-\frac{A'}{A}\right)\right\}\right.\\\nonumber
&\left.\frac{1}{A^2}+\left(\frac{\dot{A}}{A}f'_R-\dot{f'_R}
+\frac{A'}{A}\dot{f_R}\right)\left(\frac{4A'}{A}+\frac{C'}{C}
+\frac{B'}{B}\right)\frac{1}{A^4}\right.\\\label{Ad0}
&+\left.\frac{2}{A^2}\left\{-\left(\frac{A'}{A}
-\frac{C'}{C}\right)\frac{f'_R}{B^2}+\frac{\ddot{f_R}}{A^2}
-\frac{\dot{f_R}}{A^2}\left(\frac{\dot{C}}{C}+\frac{\dot{A}}{A}
\right)\right\}\frac{\dot{C}}{C}\right],\\\nonumber
&D_1=\frac{1}{\kappa}\left[\left\{\frac{1}{A^4}\left(\frac{A'}{A}
\dot{f_R}-\dot{f'_R}+\frac{\dot{A}}{A}f'_R\right)\right\}_{,0}
+\frac{1}{A^2}\left\{\frac{\ddot{f_R}}{A^2}-\frac{f'_R}{B^2}\right.\right.\\\nonumber
&\left.\left.\times\left(\frac{B'}{B}+\frac{A'}{A}+\frac{C'}{C}
\right)+\frac{f-Rf_R}{2}-\frac{\dot{f_R}}{A^2}\left(\frac{\dot{B}}{B}-\frac{\dot{A}}{A}
+\frac{\dot{C}}{C}\right)\right\}_{,1}\right.\\\nonumber
&+\left.\frac{1}{A^2}\left\{\frac{
f''_R}{A^2}-\left(\frac{A'}{A}+\frac{B'}{B}\right)\frac{f'_R}{B^2}+\left(
\frac{\dot{B}}{B}-\frac{\dot{A}}{A}\right)\frac{\dot{f_R}}{A^2}
\right\}\frac{B'}{B}\right.\\\nonumber
&\left.+\left\{\frac{f''_R}{A^2}+\frac{\ddot{f_R}}
{A^2}-2\frac{A'f'_R}{A^3}-2\frac{\dot{A}\dot{f_R}}{A^3}\right\}\frac{A'}{A^3}+\frac{1}{A^2}\left\{\frac{f''_R}{A^2}
-\frac{f'_R}{A^2}\right. \right.\\\nonumber
&\times\left.\left.\left(\frac{A'}{A}+\frac{C'}{C}\right)-\frac{
\dot{f_R}}{A^2}\left(\frac{\dot{C}}{C}-\frac{\dot{A}}{A}\right)
\right\}\frac{C'}{C}-\frac{1}{A^4}\left(\dot{f'_R}-\frac{\dot{A}}{A}f'_R\right.\right.\\\label{Ad1}
&-\left.\left.
-\frac{A'}{A}\dot{f_R}\right)\left(\frac{\dot{B}}{B}+\frac{\dot{C}}{C}
+\frac{4\dot{A}}{A}\right)\right].
\end{align}
In view of $f(R)$ model, the modified versions of structure scalars
are
\begin{align}\label{A3}
&Y_T^{(G)}=\frac{\kappa(1+\gamma{R^n})^2\left(\mu+P_z+P_r
+P_\phi\right)}{2nR^{n-1}[(1+\gamma{R^n})(2\alpha{R}^n-\beta)-\gamma{R^n}(\alpha
R^n-\beta)]}, \\\label{A4} &
Y_T^{(C)}=\frac{2\kappa(1+\gamma{R^n})^2\pi{E}^2}{nR^{n-1}B^2
[(1+\gamma{R^n})(2\alpha{R}^n-\beta)-\gamma{R^n}(\alpha
R^n-\beta)]}=Y_S^{(C)}=-Y_K^{(C)},\\\nonumber &
Y_T^{(D)}=\frac{(1+\gamma{R^n})^2}{2nR^{n-1}[(1+\gamma{R^n})(2\alpha{R}^n-\beta)-\gamma{R^n}(\alpha
R^n-\beta)]}\left[\frac{{\alpha}R^{2n}(1-2n)}{2(1+\gamma{R^n})}\right.\\\label{A5}
&\left.-\frac{{\beta}R^n(1-n)}{2(1+\gamma{R^n})}+\frac{n\gamma{R}^{2n}(\alpha{R^n}-\beta)}
{2(1+\gamma{R^n})^2}+\frac{1}{A^2}\left(\psi_{tt}+\psi_{rr}\right)
+\frac{\psi_{zz}}{B^2} +\frac{\psi_{\phi\phi}}{C^2}\right]
,\\\label{A6}
&Y_S^{(G)}=\mathbb{E}_S-\frac{\kappa(1+\gamma{R^n})^2\left(P_z-P_r
\right)}{2nR^{n-1}[(1+\gamma{R^n})(2\alpha{R}^n-\beta)-\gamma{R^n}(\alpha
R^n-\beta)]},\\\label{A7}
&Y_S^{(D)}=\frac{\kappa(1+\gamma{R^n})^2}{2nR^{n-1}[(1+\gamma{R^n})
(2\alpha{R}^n-\beta)-\gamma{R^n}(\alpha
R^n-\beta)]}\left(\frac{\psi_{rr}}{A^2}-\frac{\psi_{zz}}{B^2}\right),\\\label{A8}
&Y_K^{(G)}=\mathbb{E}_K-\frac{\kappa(1+\gamma{R^n})^2\left(P_\phi
-P_r\right)}{2nR^{n-1}[(1+\gamma{R^n})(2\alpha{R}^n-\beta)-\gamma{R^n}(\alpha
R^n-\beta)]}\\\label{A9} &
Y_K^{(D)}=\frac{\kappa(1+\gamma{R^n})^2}{2nR^{n-1}[(1+\gamma{R^n})
(2\alpha{R}^n-\beta)-\gamma{R^n}(\alpha
R^n-\beta)]}\left(\frac{\psi_{rr}}{A^2}-\frac{\psi_{zz}}{C^2}\right),\\\label{A10}
&X_S^{(G)}=-\mathbb{E}_S-\frac{\kappa(1+\gamma{R^n})^2\left(P_z-P_r\right)}
{2nR^{n-1}[(1+\gamma{R^n})(2\alpha{R}^n-\beta)-\gamma{R^n}(\alpha
R^n-\beta)]},\\\label{A11}
&X_S^{(C)}=-\frac{2\pi(1+\gamma{R^n})^2\kappa{E}^2}{nB^2R^{n-1}
[(1+\gamma{R^n})(2\alpha{R}^n-\beta)-\gamma{R^n}(\alpha
R^n-\beta)]}=X_K^{(C)}=-X_T^{(C)},\\\label{A12}
&X_S^{(D)}=\frac{\kappa(1+\gamma{R^n})^2}{2nR^{n-1}[(1+\gamma{R^n})
(2\alpha{R}^n-\beta)-\gamma{R^n}(\alpha
R^n-\beta)]}\left(\frac{\psi_{rr}}{A^2}-\frac{\psi_{zz}}{B^2}\right),\\\label{A13}
&X_K^{(G)}=-\mathbb{E}_K-\frac{\kappa(1+\gamma{R^n})^2\left(
P_\phi-P_r\right)}{2nR^{n-1}[(1+\gamma{R^n})(2\alpha{R}^n-\beta)-\gamma{R^n}(\alpha
R^n-\beta)]}\\\label{A14}
&X_K^{(D)}=-\frac{\kappa(1+\gamma{R^n})^2}{2nR^{n-1}[(1
+\gamma{R^n})(2\alpha{R}^n-\beta)-\gamma{R^n}(\alpha
R^n-\beta)]}\left(\frac{\psi_{\phi\phi}}{C^2}-\frac{\psi_{rr}}
{A^2}\right),\\\label{A15} &X_T^{(G)}
=\frac{\kappa(1+\gamma{R^n})^2(\mu)}{nR^{n-1}[(1+\gamma{R^n})
(2\alpha{R}^n-\beta)-\gamma{R^n}(\alpha R^n-\beta)]},\\\label{A16}
&X_T^{(D)}=\frac{\kappa(1+\gamma{R^n})^2\psi_{tt}}{2nA^2R^{n-1}
[(1+\gamma{R^n})(2\alpha{R}^n-\beta)-\gamma{R^n}(\alpha
R^n-\beta)]},\\\label{A17}
&Z_q^{(G)}=\frac{\kappa(1+\gamma{R^n})^2q}{2nR^{n-1}[
(1+\gamma{R^n})(2\alpha{R}^n-\beta)-\gamma{R^n}(\alpha
R^n-\beta)]},\quad Z_q^{(C)}=0,\\\label{A18}
&Z_q^{(D)}=-\frac{\kappa(1+\gamma{R^n})^2\psi_{tr}}{2nA^2
R^{n-1}[(1+\gamma{R^n})(2\alpha{R}^n-\beta)-\gamma{R^n}(\alpha
R^n-\beta)]}, \quad Z_H=2H.
\end{align}

\bsp

\label{lastpage}

\end{document}